 \documentclass[final,3p,times,twocolumn]{elsarticle}

\usepackage{graphicx}
\usepackage{amssymb}
\usepackage{amsmath}
\usepackage{nccmath}
\usepackage{mathtools}
\usepackage{float}
\usepackage{microtype}
\usepackage{tabularx}
\usepackage{multirow}
\usepackage{adjustbox}
\usepackage{makecell}
\usepackage{mhchem}
\biboptions{sort&compress}

\begin{document}

\begin{frontmatter}


\title{A study of the influence of plasma-molecule interactions on particle balance during detachment}



%\author{K. Verhaegh$^{1,2,3}$, B. Lipschultz$^2$, B.P. Duval$^3$, A. Fil$^{2,1}$, D.S. Gahle$^{4,1}$, J.R. Harrison$^1$, D. Moulton$^1$, A. Perek$^5$, A. Smolders$^3$, M. Wensing$^3$, C. Bowman$^2$, F. Federici$^2$, O. F\'{e}vrier$^3$, C. Theiler$^3$}
%\address{$^1$ Culham Centre for Fusion Energy, Culham, United Kingdom} 
%\address{$^2$ York Plasma Institute, University of York, United Kingdom}
%\address{$^3$ Swiss Plasma Centre, \'{E}cole Polytechnique F\'{e}d\'{e}rale de Lausanne, Lausanne, Switzerland}
%\address{$^4$ SUPA, University of Strathclyde, Glasgow, United Kingdom}
%\address{$^5$ DIFFER, Eindhoven, The Netherlands}
\author[CCFE,York,EPFL]{Kevin Verhaegh}
\ead{kevin.verhaegh@ukaea.uk}
\author[York]{Bruce Lipschultz}
\author[CCFE]{James Harrison}
\author[EPFL]{Basil Duval}
\author[York]{Chris Bowman}
\author[York,CCFE]{Alexandre Fil}
%\author[EPFL]{Olivier F\'{e}vrier}
\author[SUPA,CCFE]{Daljeet Singh Gahle}
\author[CCFE]{David Moulton}
\author[York,CCFE]{Omkar Myatra}
\author[DIFFER]{Artur Perek}
\author[EPFL]{Christian Theiler}
\author[EPFL]{Mirko Wensing}
\author[MST1]{MST1 team}
\author[TCV]{TCV team}
%\author[York]{Fabio Federici}
%

\address[CCFE]{Culham Centre for Fusion Energy, Culham, United Kingdom}
\address[York]{York Plasma Institute, University of York, United Kingdom}
\address[EPFL]{Swiss Plasma Centre, \'{E}cole Polytechnique F\'{e}d\'{e}rale de Lausanne, Lausanne, Switzerland}
\address[SUPA]{SUPA, University of Strathclyde, Glasgow, United Kingdom}
\address[DIFFER]{DIFFER, Eindhoven, The Netherlands}
\address[TCV]{See author list of "S. Coda et al 2019 Nucl. Fusion 59 112023"}
\address[MST1]{See author list of "B. Labit et al 2019 Nucl. Fusion 59 086020"}

\begin{abstract}
%In this work we provide experimental insights into the impact of plasma-molecule interactions on divertor detachment by applying new spectroscopic analysis techniques to the hydrogen Balmer line series to investigate how both atom and plasma-molecule interactions impact particle balance.

%Our analysis on a representative L-mode TCV density ramp discharge indicates that Molecular Activated Recombination ion sinks from $H_2^+$ and/or $H^-$ are significantly (a factor $\sim$ 5) larger than Electron-Ion Recombination for TCV and are a significant contributor to the observed reduction in the outer divertor ion target flux. Such MAR reactions lead to the formation of excited atoms which contribute significantly to the measured hydrogenic line emission spectra and thus hydrogenic radiation. Such contributions should be accounted for when analysing the hydrogenic line series for inferring ionisation rates. %As detachment proceeds both atomic Balmer line emission from electron-impact excitation (which is an indicator for ionisation) and Fulcher band emission (from excited molecules from plasma-molecule \emph{collisions}) detach from the target while Balmer line emission from excited atoms after plasma-molecule \emph{reactions} remain peaked near the target and extend until the electron-impact excitation emission region.

In this work we provide experimental insights into the impact of plasma-molecule interactions on the target ion flux decrease during divertor detachment achieved through a core density ramp in the TCV tokamak. Our improved analysis of the hydrogen Balmer series shows that plasma-molecule processes are strongly contributing to the Balmer series intensities and substantially alter the divertor detachment particle balance.

We find that Molecular Activated Recombination (MAR) ion sinks from $H_2^+$ (and possibly $H^-$) are a factor $\sim$ 5 larger than Electron-Ion Recombination (EIR) and are a significant contributor to the observed reduction in the outer divertor ion target flux. Molecular Activated Ionisation (MAI) appears to be substantial during the detachment onset, but further research is required into its magnitude given its uncertainty.  

Plasma-molecule interactions enhance the Balmer line series emission strongly near the target as detachment proceeds. This indicates enhancements of the Lyman series, potentially affecting power balance in the divertor. As those enhancements vary spatially in the divertor and are different for different transitions, they are expected to result in a separation of the $Ly\beta$ and $Ly\alpha$ emission regions. This may have implications for the treatment and diagnosis of divertor opacity.

The demonstrated enhancement of the Balmer series through plasma-molecule processes potentially poses a challenge to using the Balmer series for understanding and diagnosing detachment based only on atom-plasma processes.

\end{abstract}


\end{frontmatter}

%\linenumbers

\section{Introduction}
\label{ch:introduction}

Divertor detachment is expected to be a crucial aspect for handling the power exhaust of future fusion devices, such as ITER and DEMO \cite{Pitts2013}. During detachment, a range of atomic and molecular processes result in simultaneous power, particle and momentum losses from the plasma to neutral species or to photons (e.g. radiative power loss). This results in a simultaneous reduction of the target plasma temperature ($T_t$) and the ion target flux $\Gamma_t$. Such changes in the plasma facilitate large reductions in the target heat flux ($q_t$) as shown in equation \ref{eq:qt} where $\gamma$ is the sheath transmission coefficient and $\epsilon$ is the surface recombination energy that is deposited when an $H^+$ \footnote{In this work we use $H$ for hydrogen as the available reaction rates/emission coefficients are only available for hydrogen. The discharge discussed in this work is, however, a deuterium discharge. The impact of such assumptions are further discussed in section \ref{ch:analysis} following \cite{Verhaegh2020}.} ion converts to an atom (13.6 eV) and afterwards to a molecule ($+2.2$ eV) \cite{Loarte2007}.

\begin{equation}
    q_t = \Gamma_t (\gamma T_t + \epsilon)
    \label{eq:qt}
\end{equation}

This simultaneous reduction of the ion target flux ($\Gamma_t$) and the target temperature requires target pressure ($p_t$) loss according to the sheath-target conditions (equation \ref{eq:sheath}). That target pressure loss can be facilitated through volumetric momentum losses \cite{Goetz1996,LaBombard1997,Pitcher1997,Labombard1995,Stangeby2000,Stangeby2018} and may involve an upstream pressure loss as indicated in previous research on TCV \cite{Verhaegh2019}. 

\begin{equation}
    \Gamma_t \propto p_t / T_t^{1/2}
    \label{eq:sheath}
\end{equation}

%The importance of detachment for reducing the target heat flux is illustrated by asking how much power loss can occur during attached plasmas where $p_t$ is constant. As $T_t$ drops, $\Gamma_t$ (equation \ref{eq:sheath}) increases while $q_t$ drops - but not very much: assuming $T_t$ drops from 100 eV to 1 eV, one would have a reduction of $q_t$ of only a factor $2.5$ assuming $\gamma = 5$ and $\epsilon = 15.8$ eV. To achieve higher heat flux losses, one would require a simultaneous reduction of the ion target flux, the target temperature and the target pressure: detachment.

A reduction of the ion flux requires either a reduction of the ion source or an increase in the ion sink according to particle balance (equation \ref{eq:PartBal}). Here, $\Gamma_i$ ($\Gamma_r$) is the divertor ion source (sink) and $\Gamma_u$ is a net ion flow from upstream towards (positive) / away from (negative) the target. $\Gamma_t$ is generally thought to be much larger than $\Gamma_u$: $\Gamma_t \ll \Gamma_u \rightarrow \Gamma_t \approx \Gamma_i -\Gamma_r$: high recycling conditions  (\cite{Lipschultz1999,Krasheninnikov2017,Pshenov2017,Stangeby2018,Verhaegh2019} and figure \ref{fig:HaMeasurement}). %In these conditions, a reduction in the ion target flux requires either a reduction of the divertor ion source $\Gamma_i$ and/or an increase of the divertor ion sink $\Gamma_r$. 

\begin{equation}
    \Gamma_t = \Gamma_i - \Gamma_r + \Gamma_u
    \label{eq:PartBal}
\end{equation}

 \begin{figure}[H]
    \centering
    \includegraphics[width=\linewidth]{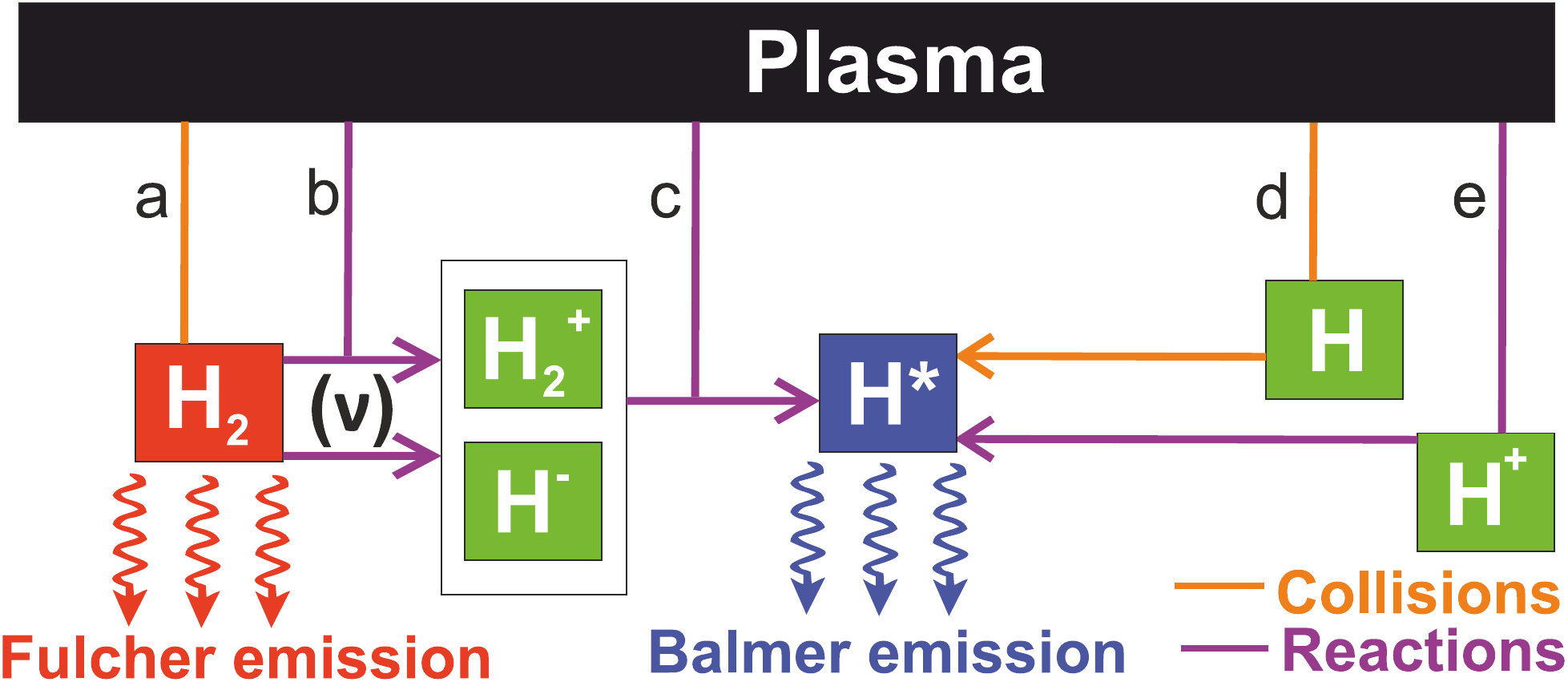}
    \caption{Schematic overview of important plasma-molecule and plasma-atom collisions (in orange) and reactions (in magenta) discussed in this work. a) Collisions between the plasma and $H_2$ excite the molecule rovibronically; b) Reactions between the plasma and $H_2 (\nu)$ result in the formation of $H_2^+$ and $H^-$; c) Reactions between the plasma and $H_2^+$ (and possibly $H^-$), which can result in excited neutral atoms ($H*$) and Balmer line emission; d) Collisions between electrons and $H$ excite $H*$, resulting in Balmer line emission; e) Electron-ion recombination reactions with $H^+$ resulting in excited $H*$ and Balmer line emission.}
    \label{fig:FulcherBalmerSchem}
\end{figure}

The divertor ion source ($\Gamma_i$) is strongly connected to power balance as each ionisation event requires a certain amount of energy ($E_{ion}$) \cite{Lipschultz1999,Krasheninnikov2017,Pshenov2017,Stangeby2018,Verhaegh2019}. The ionisation source ($\Gamma_i$) becomes limited by the power flux flowing into the recycling/ionisation region ($q_{recl}$) if the power flux loss due to ionisation ($\Gamma_i E_{ion}$) becomes comparable (within a factor of two \cite{Verhaegh2019}) to $q_{recl}$. Impurity radiation is often (one of) the main power dissipation processes reducing the power crossing the SOL ($q_{SOL}$) before the power enters the recycling region ($q_{recl}$). Particle, power and momentum balance are thus all connected through equations \ref{eq:qt}, \ref{eq:sheath} and \ref{eq:PartBal} (where $q_t = q_{recl} - E_{ion} \Gamma_i$). All three balances play an important role in detachment \cite{Stangeby2017,Verhaegh2019,Stangeby2000}. Most studies of detachment have focussed on the target ion current in detachment and how plasma-atom interactions affect power, momentum and particle balance, leading $\Gamma_t$ to drop.

%Plasma-molecule \emph{collisions} excite the molecules rovibronically,  resulting in vibrationally "($/nu$)" excited molecules and Fulcher band emission through electronically excited $H_2$ states. Vibrational excitation promotes the creation of $H^-$ and $H_2^+$ (the latter only through molecular charge exchange) resulting in  Balmer line emission through \emph{reactions} between the plasma and $H_2^+$ and $H^-$. Plasma-atom interactions (electron-impact excitation and EIR) also result in Balmer line emission
 
Modifications in those three balances are driven by plasma-atom \cite{Verhaegh2019,Lipschultz1999,Lomanowski2019} and plasma-molecule interactions.  To investigate the impact of plasma-molecule interactions further, we must distinguish between two different 'types' of plasma-molecule interactions: ('elastic') \emph{collisions} between the plasma and $H_2$ and \emph{reactions} from the plasma chemistry of $H_2$. A schematic overview of such \emph{reactions} and \emph{collisions} are shown in figure \ref{fig:FulcherBalmerSchem}. Collisions between the plasma and $H_2$ lead to rovibronic excitation \cite{Fantz2002,Fantz2001,Sakamoto2017,Hollmann2006,Groth2019,Kukushkin2017,Wischmeier2004} (figure \ref{fig:FulcherBalmerSchem} a). 
 
Reactions between the plasma and $H_2$ leads to $H_2$ dissociation and to the formation of $H_2^+$ and $H^-$ \cite{Fantz2001,Fantz2002,Wuenderlich2016} as illustrated in \ref{fig:FulcherBalmerSchem} b. Reactions which result in the \emph{formation} (for $T_e < 4$ eV) of $H_2^+$ and $H^-$ can be greatly enhanced by the population of higher vibrationally excited states ($\nu)$ which are populated through plasma-molecule \emph{collisions} \cite{Fantz2002} (figure \ref{fig:FulcherBalmerSchem}a). $H_2^+$ and $H^-$ react with the plasma (figure \ref{fig:FulcherBalmerSchem}c), \emph{breaking down} $H_2^+$ and $H^-$, resulting in excited ($*$) neutrals which emit atomic hydrogen line emission \cite{Wuenderlich2016,Wunderlich2020,Verhaegh2020,Sakamoto2017} - see figure \ref{fig:FulcherBalmerSchem}c. The cross-sections for the reactions which form $H_2^+$ and $H^-$ can have strong isotope dependencies which are still being debated in literature \cite{Kukushkin2016,Krishnakumar2011}. In particular the dissociative attachment ($H_2 + e^- \rightarrow H^- + H$) cross-section is thought to be strongly diminished for deuterium (\ce{^{2}_{1} H+}) and more so for tritium (\ce{^{3}_{1} H+}) compared to protium (\ce{^{1}_{1} H+}) \cite{Krishnakumar2011}. %: $H_2^+$ is formed by Molecular charge exchange ($H^+ + H_2 \rightarrow H_2^+ + H$); dissociative attachment of $H_2$ ($H_2 + e^- \rightarrow H^- + H$) results in the formation of $H^-$. These "creation" reactions for $H_2^+$ and $H^-$ are summarised in table \ref{tab:ReactionRates}.

Most experimental investigations into the impact of molecules on plasma-edge physics in tokamaks have utilised $H_2$ Fulcher band (590-640 nm) emission measurements \cite{Hollmann2006,Fantz2002} which arise from electronic excitation of $H_2$ due to plasma-molecule \emph{collisions} (step a in figure \ref{fig:FulcherBalmerSchem}). Those measurements directly show that that the plasma is interacting with the molecules resulting in rovibronic (meaning electronically - resulting in Fulcher band emission, vibrationally - $H_2 (\nu)$ and rotationally) excitation \cite{Fantz2001,Fantz2002,Hollmann2006}. That information, combined with model or simulation results, has been used in studies to infer information on \emph{reactions} between the plasma and molecules \cite{Fantz2002,Hollmann2006}.

Although it was suspected from DIII-D and JET studies \cite{Lomanowski2020,Hollmann2006} that the $H\alpha$ emission in the divertor may be enhanced by plasma-molecule interactions, the impact of excited atoms from plasma-molecule interactions on the atomic spectra has not yet been studied quantitatively in tokamak divertors. %That is the subject of this paper. We achieve that by employing a new technique to analyse the Balmer line series to extract information about plasma-molecule interactions \cite{Verhaegh2020} to experimental data from the TCV tokamak, which we utilise to estimate the impact of plasma-molecule interactions on divertor particle balance. 

%Dissociative attachment as well as molecular charge exchange (step b in figure \ref{fig:FulcherBalmerSchem}, which involves the "creation" reactions for $H_2^+$ and $H^-$ listed in table \ref{tab:ReactionRates}) can be greatly enhanced by the population of higher vibrationally excited states ($\nu$) which can be populated through plasma-molecule \emph{collisions} \cite{Fantz2002} (figure \ref{fig:FulcherBalmerSchem}a).  (see the rates in table \ref{tab:ReactionRates}). For instance: $H_2^+$ can undergo dissociative recombination with an electron $H_2^+ + e^- \rightarrow H + H^*$; and $H^-$ can undergo mutual neutralisation with an ion $H^- + H^+ \rightarrow H + H^*$. Many of such reactions ultimately result in neutral (often excited $*$) atoms, which emit atomic hydrogen line emission through the various series such as Balmer and Lyman \cite{Wuenderlich2016,Wunderlich2020,Verhaegh2020,Sakamoto2017} - see figure \ref{fig:FulcherBalmerSchem}c. 

\subsection{The effect of plasma-molecule interactions on particle, energy and momentum balance}

There are multiple chains of plasma-molecule reactions involving $H_2$, $H_2^+$ and $H^-$ that 'effectively' ionise neutrals and recombine ions; a summary can be found in \cite{Verhaegh2020}. One example is molecular charge exchange ($H_2 + H^+ \rightarrow H_2^+ + H$) resulting in the formation of $H_2^+$ which dissociatively recombines with an electron: $H_2^+ + e^- \rightarrow H + H$. When comparing the inputs and outputs of those reactions, we observe that an ion was effectively recombined: Molecular Activated Recombination - MAR. Alternatively at higher temperatures, molecular reactions can activate the ionisation of a plasma neutral - Molecular Activated Ionisation - MAI.

Both plasma-molecule collisions and reactions impact the plasma power balance. Collisions between the plasma and the molecules transfer kinetic energy from the plasma to the molecular cloud \cite{Park2018,Myatra,Smolders}. Although emission (and thus radiation) from molecular ($H_2$) bands occurs, the radiative losses from such processes are experimentally estimated to be insignificant - in agreement with EDGE2D-Eirene modelling \cite{Groth2019}. Radiative energy loss also occurs from excited atoms formed after plasma-molecule reactions \cite{Wuenderlich2016,Wunderlich2020,Verhaegh2020} (step c in figure \ref{fig:FulcherBalmerSchem}).

%Both plasma-molecule collisions and reactions can impact the momentum balance of the plasma. 
Plasma-molecule collisions transfer momentum from the plasma to the molecules, effectively acting as a momentum sink \cite{Park2018,Moulton2018,Myatra,Smolders,Stangeby2017}. Apart from collisions, the molecular charge exchange reaction ($H_2 + H^+ \rightarrow H_2^+ + H$) also results in momentum losses \cite{Moulton2018,Myatra}. %\textid{(Comment - not clear here. It seems there is a general statement that there are many plasma moledcule collisions that result in plasma momentum loss. That is followed by a spcific process?)}.

\subsection{The scope of this paper}

In this work we employ a new technique \cite{Verhaegh2020} to experimental Balmer spectra from the TCV tokamak to extract information about plasma-molecule interactions. Our results indicate plasma-molecule interactions strongly increase hydrogenic line emission and modify particle balance significantly during detachment. MAI starts to contribute to the ion target flux at around the detachment onset. For the TCV case studied, MAR is $\sim$ 5 times larger than the EIR ion sink (atomic processes). The final result is that the inclusion of plasma sources and sinks due to plasma-molecule processes significantly alters the picture of particle balance derived from plasma-atom processes alone. %Our analysis separates the molecular contributions to the Balmer line emission spectra ($n=3-6$) which are then used to infer the magnitude and location of MAI/MAR ion sources/sinks accounting for the possible $H_2^+$ and $H^-$ reactions. Our results indicate that plasma-molecule \emph{reactions} involving $H_2^+$ (and/or $H^-$) lead to significant levels of both MAR and MAI.

The Balmer line emission enhancement during detachment attributed to plasma-molecule interactions: 1) is strongest near the target while the $H_2$ Fulcher emission region tracks the ionisation region; suggesting that different plasma-molecule interactions occur at different locations of the TCV divertor; 2) are indicative of increases of the Lyman series, which could potentially affect power balance in the divertor; 3) could have implications for divertor opacity effects as it can result in a spatial separation between the $Ly\alpha$ and $Ly\beta$ emission regions.

\section{Spectral analysis techniques of inferring information on plasma-molecule interaction from the Balmer spectra}
\label{ch:analysis}

We have used the analysis technique 'Balmer Spectroscopy of Plasma-Molecular Interactions' - 'BaSPMI' \cite{Verhaegh2020} to investigate the spectroscopic data in this work. For simplicity we define Balmer line emission from electron-impact excitation of $H$ and EIR of $H^+$ as \emph{"atomic"} contributions to the Balmer line emission while we define Balmer line emission arising from excited atoms after plasma interactions with $H_2, H_2^+$ and $H^-$ as contributions to the Balmer line emission arising from \emph{"plasma-molecule interactions"}. The total atomic line emission is then the sum of the \emph{"atomic"} ($H, H^+$) and \emph{"plasma-molecule interaction"} ($H_2, H_2^+, H^-$) contributions \cite{Verhaegh2020}.  BaSPMI uses chordal-integrated brightness measurements of $H\alpha, H\beta$, as well as two other medium-n Balmer lines (e.g. n=5-7) together with Stark inferred electron densities; and uses data from ADAS \cite{OMullane,Summers2006}, YACORA (On the Web) \cite{Wuenderlich2016,Wunderlich2020} as well as AMJUEL \cite{Reiter2005}.

The goal of BaSPMI is "to quantify the contribution of plasma-molecule interactions to the Balmer lines and use this to provide quantitative estimates of the influence of molecules on power losses, particle (ion) sources/sinks and Balmer line emission" \cite{Verhaegh2020}. This builds upon previous atomic analysis technique developed by the authors in \cite{Verhaegh2019a} and works on the principle that certain plasma-molecule \emph{reactions} result in \emph{excited atoms} emitting \emph{atomic line emission}, which is more dominant for lower-n Balmer lines ($H\alpha, H\beta$) than higher-n ones. 

The profile of the various emission contributions vary continuously along the line of sight. However, the analysis technique (see \cite{Verhaegh2020}) simplifies the emission profiles as a 'dual slab' model (with a hot and cold temperature) along the line of sight. That approach has been verified using SOLPS-ITER simulations for both TCV (density scan) and MAST-U (density and $N_2$ puffing scan) using 'synthetic testing'. Although there is a strong spatial separation of the various emission and reaction regions in those simulations, the analysis of the \emph{chordal-integrated brightnesses} still provides similar estimates of the \emph{chordal-integrated atomic and molecular ion sources/sinks} as is obtained when summing those sources/sinks from the simulation directly along the line of sight. The 'synthetic testing', employed in \cite{Verhaegh2020}, uses data tables from \cite{Kukushkin2016} for the $H_2 + H^+ \rightarrow H_2^+ + H$ rate (for deuterium) and default AMJUEL data tables (protium) for other rates related to $H_2^+$ and $H^-$ to post-process the SOLPS-ITER simulations to obtain the $H_2^+$ and $H^-$ densities. Similar agreement between the analysis result and the result directly obtained from the simulation is obtained when either the AMJUEL base rate (protium) or the 'remapped ($T_e$/2)' AMJUEL rate (Eirene default for deuterium) for $H_2^+ + H^+ \rightarrow H_2^+ + H$ although similar results are obtained with different rates (AMJUEL base rates (protium) and 'remapped' ($T_e/2$) rates (deuterium) \cite{Verhaegh2020}). Furthermore, the emission contributions from $H_2^+$ and $H^-$ have been turned off individually in the synthetic testing in to verify that the analysis correctly identifies their presence \cite{Verhaegh2020}.

Although the cross-sections for \emph{forming} $H_2^+$ and $H^-$ have strong isotope dependencies (see section \ref{ch:introduction}), BaSPMI does not use these rates \cite{Verhaegh2020} and instead detects the excited neutrals arising from reactions \emph{breaking down} $H_2^+$ and $H^-$. However, BaSMPI does rely on population coefficients from Yacora (on the Web) \cite{Wuenderlich2016,Wunderlich2020} and MAR/MAI rates from AMJUEL \cite{Reiter2008}, which are available for just protium. Nevertheless, the analysed discharge is a deuterium discharge. We provide the analysis in this work with this caveat and increased availability of data for deuterium would benefit this analysis \cite{Verhaegh2020}. %Nevertheless, we think that the presented results are applicable to both deuterium and protium plasmas.  

Although the cross-section for dissociatve attachment are thought to be strongly reduced for deuterium discharges \cite{Krishnakumar2011}, we keep the hydrogenic line emission arising from plasma interactions with $H^-$ as a free parameter in this work. Removing that degree of freedom does not change any of the presented results beyond their uncertainty margins as the emission attributed to $H^-$ will now be attributed to $H_2^+$ instead and the MAR/$H\alpha$ photon ratio for $H^-$ and $H_2^+$ are similar within experimental uncertainties \cite{Verhaegh2020}. We have bundled the contributions of $H_2^+$ and $H^-$ together in terms of their Balmer line emission and MAR/MAI ion sinks/sources throughout this work as distinguishing between those two is beyond the scope of this paper.

% Future work will focus more on the differences between Balmer line emission arising from reactions with $H^-$ and $H_2^+$ as well as the two different MAR ion sinks.

%In this work we will focus on the combined emission due to plasma-molecule interaction for the various Balmer line measurements as well as the combined MAR and MAI rates to influence the impact on particle balance. A more detailed investigation also including power loss estimates as well as separating contributions from $H^-$ and $H_2^+$ will be presented in future work \cite{}.

\section{Experimental results on TCV}
\label{ch:results}

\begin{figure}[H]
    \centering
    \includegraphics[width=\linewidth]{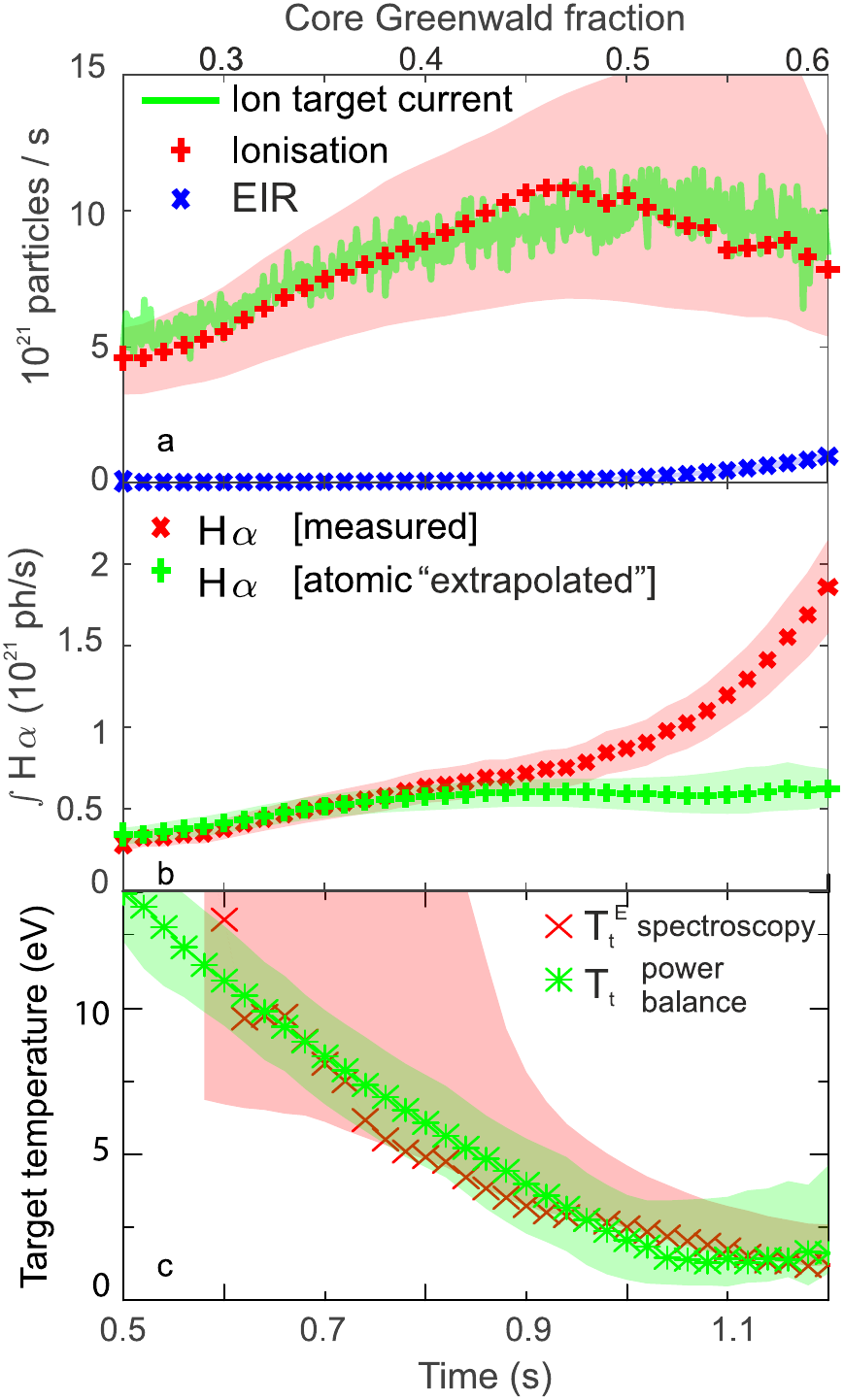}
    \caption{Inferred particle balance only considering plasma-atom interactions. a) Particle balance (ion target current) shown together with estimated atomic ion source, electron-ion recombination (EIR) sink. b) Measured total $H\alpha$ in the divertor together with the estimated atomic part of the $H\alpha$ emission. c) Target temperature estimates (based on the spectroscopically inferred excitation temperature - $T_e^E$ and a temperature estimate obtained from power balance, adopted from \cite{Verhaegh2019}.}    
    \label{fig:HaMeasurement}
\end{figure}

%In the following we apply the above analysis chain (BaSPMI) to a representative discharge in TCV to demonstrate the role of plasma-molecule interactions on particle balance.
%In this work we will highlight the possible impacts of plasma-molecule interactions on particle balance and the evolution of both Balmer line emission as well as $H_2$ Fulcher band emission associated with plasma-molecule interactions during detachment along the divertor leg using a single typical reference deuterium density ramp discharge from TCV. 
The TCV discharge used for this study (\#56567) has previously been analysed in detail from the point of view of atomic interactions only \cite{Verhaegh2019}. As shown in Fig. \ref{fig:HaMeasurement}d, \#56567 has a single null divertor shape with spectroscopy lines of sight intersecting the outer divertor leg along most of its length. \#56567 is an L-mode Deuterium plasma with $I_p = 340 kA$ discharge in reversed field (i.e. ion grad-B drift away from the primary x-point) without additional impurity seeding (although intrinsic carbon impurities are present and are an important power loss process \cite{Verhaegh2019}). This discharge has been repeated multiple times with different spectroscopy settings to obtain sufficient spectroscopic coverage for BaSPMI as well as to obtain Fulcher band measurements. The reproducibility of these discharges is adequate for this \cite{Verhaegh2019}.%and has been found to be adequate for this purpose.

First we investigate particle balance of this discharge (\ref{fig:HaMeasurement}) based on a spectroscopic analysis \cite{Verhaegh2019a} which assigns all medium-n Balmer line emission to  \emph{atomic interactions only}. Under TCV attached conditions we observe a \emph{linear} increase of the ion target current (figure \ref{fig:HaMeasurement} a). Such a linear increase is predicted by analytical models, due partially to the reduction of upstream temperature during the ramp of the upstream density \cite{Verhaegh2019}. The total ion target current, $\Gamma_t$, stops rising linearly at around a Greenwald fraction of 0.33 (detachment onset) and rolls-over around a Greenwald fraction of 0.4. We observe that the atomic electron-ion recombination sink only becomes relevant at latest phase of the discharge (where the target temperature $T_t \sim 1$ eV - figure \ref{fig:HaMeasurement}c), while the ionisation rate drops before this phase at the detachment onset (where $T_t \sim 4$ eV - figure \ref{fig:HaMeasurement}c). Our atomic analysis has shown that the reduction of the ionisation source occurs as the power entering the recycling region becomes 'limited' to sustain sufficient ionisation - 'power limitation' \cite{Verhaegh2019}, which has been theorised \cite{Krasheninnikov1997,Lipschultz1999} and suspected from experiments previously \cite{Lipschultz1999,Lomanowski2019}. 

A bifurcation starts to occur between the measured and estimated "atomic extrapolated" $H\alpha$ emission at the detachment onset ($T_t \sim 4$ eV - figure \ref{fig:HaMeasurement}c), figure \ref{fig:HaMeasurement} b, and become increasingly more significant ($T_t<2$ eV - figure \ref{fig:HaMeasurement}c) as the divertor becomes colder. This is indicative of an additional source of $n=3$ excited atoms. The 'atomic' $H\alpha$ emission is based on the analysis of the  medium-n Balmer lines, assuming the higher-n Balmer lines are only populated by "atomic" interactions \cite{Verhaegh2019a}. 

While the expected $H_2$ densities from SOLPS-ITER simulations under these density/temperature conditions (e.g. detached $T_t<5$ eV plasma, $n_e \sim 10^{20} m^{-3}$) contribute less than 1\% of the measured $H\alpha$ emission \cite{Verhaegh2020}, plasma-molecule interactions involving $H_2^+$ (and possibly $H^-$) could explain the additional $H\alpha$ brightness after detachment \cite{Verhaegh2020}. Such plasma-molecule interactions lead to losses (sinks) for ions in the plasma through Molecular Activated Recombination (MAR). Other possible additional sources of $n=3$ excited atoms, such as $Ly\beta$ opacity and plasma-molecule interactions with hydrocarbons, are estimated to only increase the $H\alpha$ emission by a few percent for these TCV conditions \cite{Verhaegh2020}. Therefore, we assume that the additional $H\alpha$ emission is due to plasma-molecule interaction.

%The enhancement of $H\alpha$ after detachment is qualitatively consistent with measurements from both DIII-D \cite{Hollmann2006} and JET \cite{Lomanowski2020} where inconsistencies between the measured and expected $H\alpha$ emission was suspected, which will be discussed in more detail in section \ref{ch:ImplOtherDev}. 

\subsection{The evolution of detachment with plasma-molecule interactions}
\label{ch:DetachEvolution}

\begin{figure}[H]
    \centering
    \includegraphics[width=\linewidth]{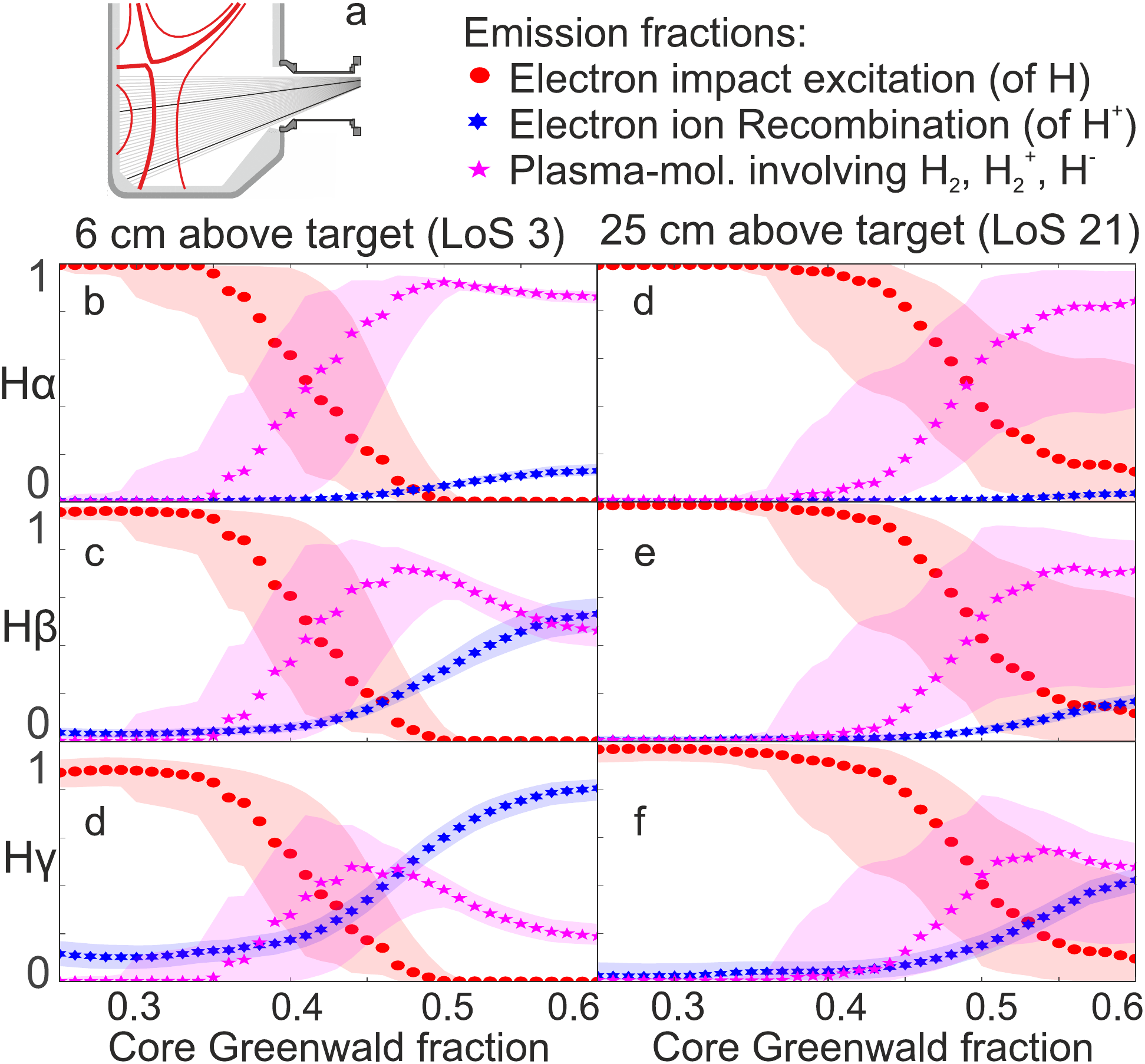}
    \caption{ a) Magnetic equilibrium of the investigated discharge with spectroscopic line of sight coverage (two chords are highlighted which are used in figure \ref{fig:SepaEmiss}. b-f) The $H\alpha, H\beta, H\gamma$ self-consistent emission fractions as function of core Greenwald fraction for two different chords (figure \ref{fig:HaMeasurement}) in terms of "atomic" contributions (electron impact excitation (of $H$) and EIR (of $H^+$) and "plasma-molecule interaction" related contributions.}
    \label{fig:SepaEmiss}
\end{figure}

In figure \ref{fig:SepaEmiss} we apply the full self-consistent BaSPMI \emph{atomic and molecular} spectroscopic analysis chain to identify the impact of plasma-molecule interactions on three measured Balmer lines measured during this discharge. Initially only electron impact excitation (of $H$) emission plays a role for all three Balmer lines (figure \ref{fig:SepaEmiss}). 

\begin{figure}[H]
    \centering
    \includegraphics[width=\linewidth]{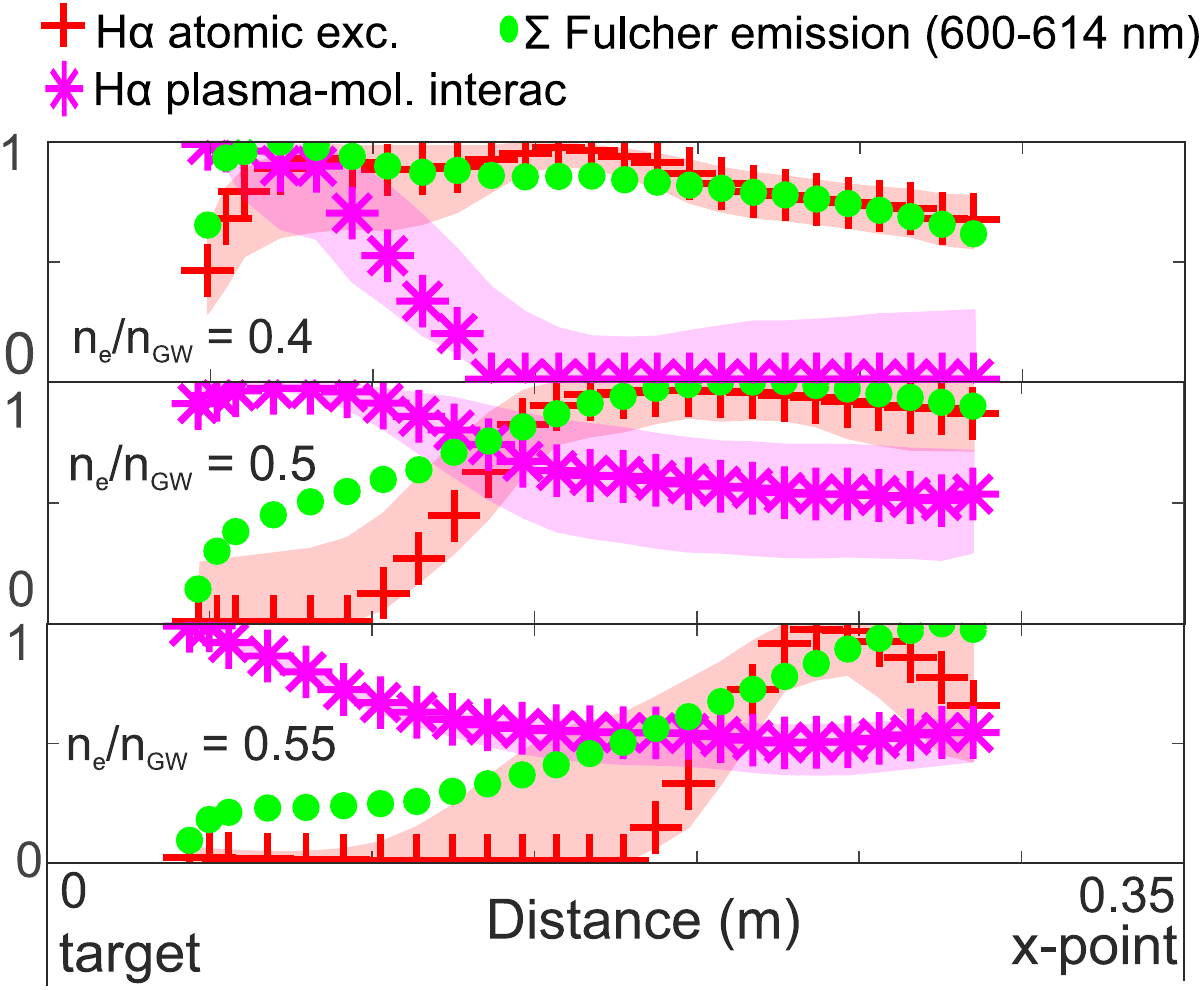}
    \caption{Spatial profiles at three different times/core Greenwald fractions ($n_e/n_{GW}$) (of the same discharge as shown throughout the paper) of normalised (to the maximum) $H\alpha$ atomic excitation emission, $H\alpha$ combined emission due to plasma molecule interactions ($H_2^+, H^-, H_2$) and summed Fulcher emission between 600 and 614 nm (which has the brightest Fulcher emission lines) with impurity emission lines removed from the spectra.}
    \label{fig:HaFulcherEmiss}
\end{figure}

As the target temperature drops, first plasma-molecule interactions and later electron-ion recombination form an increasingly larger part of the Balmer line emission (figure \ref{fig:SepaEmiss}). Near the detachment onset (Core Greenwald fraction $n_e/n_{GW} \sim 0.43$), the region of strong electron impact excitation 'detaches' from the target (see figure \ref{fig:SepaEmiss} b-d), leaving a region where enhanced Balmer line emission from $H_2^+$ (and possibly $H^-$) as well as EIR occur. This region expands as the divertor becomes colder, following the movement of the electron-impact excitation region. In the latest phases of the discharge ($n_e/n_{GW}>0.5$), "plasma-molecule interactions" are the dominant excitation process for $H\alpha$ and $H\beta$ near the target. For $H\gamma$ and $H\delta$ (not shown), the impact of plasma-molecule interactions is significant, but most of the emission is from EIR. At this phase of the discharge, the \emph{fractions} of $H\beta, H\gamma$ emission associated with plasma-molecule interactions (but not the actual \emph{brightnesses} - see figure \ref{fig:HaFulcherEmiss}) are higher 25 cm above the target than near the target due to a lack of electron-ion recombination ($T_e \sim [2.5-4]$ eV). The fact that plasma-molecule interactions are an important excitation process for $H\alpha$ and $H\beta$ emission suggests that such interactions result in significant hydrogenic radiation and thus power balance, which will be investigated in future work.

 %At the detached phase, the emission from EIR and plasma-molecule interactions increases further. %Both EIR as well as plasma-molecule reactions with $H_2^$ of those emission processes are significant for $H\beta$. 

%\subsubsection{Collisions between the plasma and $H_2$ compared to MAR/radiative loss reactions involving $H_2^+$ and $H^-$}

%Using the BaSPM analysis technique discussed in section \ref{ch:analysis} \cite{Verhaegh2020}, we obtain information from plasma-molecule \emph{reactions} predominantly with $H_2^+$ (and possibly $H^-$). However, there are also \emph{collisions} between the plasma and $H_2$, which \emph{excite the molecular cloud}. When electrons have sufficient energy for these collisions to result in excited $H_2$ electronic states and thus $H_2$ Fulcher band emission.  %losses is dominated by ionisation [see references from \emph{D.S. Gahle private communications}].
%Such collisions provide power and momentum transfer from the plasma to the molecular cloud, raising its gas temperature and increasing the densities of the higher vibrationally excited states. Vibrationally exciting $H_2$ promotes the creation of $H_2^+$ and $H^-$.

To provide some insight into the evolution of different kinds of plasma-molecule interactions during detachment, we compare the brightness profile (\emph{line-integrated} along the divertor leg and therefore also intersecting the private flux and common flux regions) of a part of the $H_2$ Fulcher band (600-614 nm) with that of the various $H\alpha$ excitation sources in figure \ref{fig:HaFulcherEmiss}. Fulcher band emission occurs when electrons have sufficient energy for plasma-molecule \emph{collisions} to result in excited electronically excited molecules. A first observation is that the Fulcher emission penetrates throughout the divertor leg in the attached phase.% This is surprising as the $H_2$ mean free paths near the target are a few centimetre according to simulations \cite{Fil2017}. The measurements thus suggest that molecules enter the divertor leg radially throughout the ionisation region, thus making it to a higher temperature region. 
%In these highly ionising conditions there is significant excitation to excited singlet states resulting in band emission such as the Fulcher bands ($A1^\Sigma_g \rightarrow X1^\Sigma_u$). As the Fulcher emission will peak where the product of the molecular density and where the $X1^\Sigma_u$ population rate is highest this suggests a significant influx of molecules \emph{from the strike point up} the divertor leg as opposed to a radial influx of molecules from \emph{around} the divertor leg. % As molecules in this attached phase are quickly dissociated near the target, this Fulcher band emission likely arises from the 'edges' of the divertor leg (see figure \ref{fig:HaMeasurement}d) from molecules forming \emph{around} the divertor leg as opposed to molecules going \emph{from the strike point up} the divertor leg.  

A second observation is that the Fulcher emission profile spatially follows the electron impact excitation (of $H$) emission profile and thus the ionisation profile. In contrast, the Balmer line emission from excited atoms after reactions with $H_2^+$ (and possibly $H^-$) occurs below the Fulcher emission region and remains peaked near the target throughout the discharge; this suggests that there is a spatial and temperature \emph{ordering} of the various molecular processes. Those two observations will be further discussed in section \ref{ch:FulcherDiscus}.
% This suggests that the ionisation region is likely similar to the region where molecules are being electronically excited, dissociated as well as ionised. %which can be expected from high energy threshold to produce excited singlet states of $H_2$ (which are comparable to the $H$ excitation energies and the source of the $H_2$ band emission [see references from \emph{D.S. Gahle private communications}]). 
%  
 %\ref{ch:FulcherDiscus}. %This is likely due to the higher molecular densities combined with a level of vibrational excitation which is still sufficient to promote $H_2^+$ and/or $H^-$ creation. 

\subsection{Particle balance with plasma-molecule interactions}

The evidence presented in the previous section suggests plasma-molecule interactions impact hydrogenic atomic line emission significantly. This has implications for particle balance in two ways: 1) When the hydrogenic Balmer spectra is attributed to be solely from "atomic" interactions (as in \cite{Verhaegh2019}), any Balmer line emission from plasma-molecule interactions will be attributed to "atomic processes"; that 'inflates' the atomic ion source/sink estimates. 2) When plasma-molecule interactions contribute to, and are properly accounted for, in the Balmer line emission, that indicates the presence of an additional ion sink through Molecular Activated Recombination (MAR) and/or ion source through Molecular Activated Ionisation (MAI). %As will be shown next, both MAR and MAI are important in the particle balance of discharge studied.

\begin{figure}[H]
    \centering
    \includegraphics[width=\linewidth]{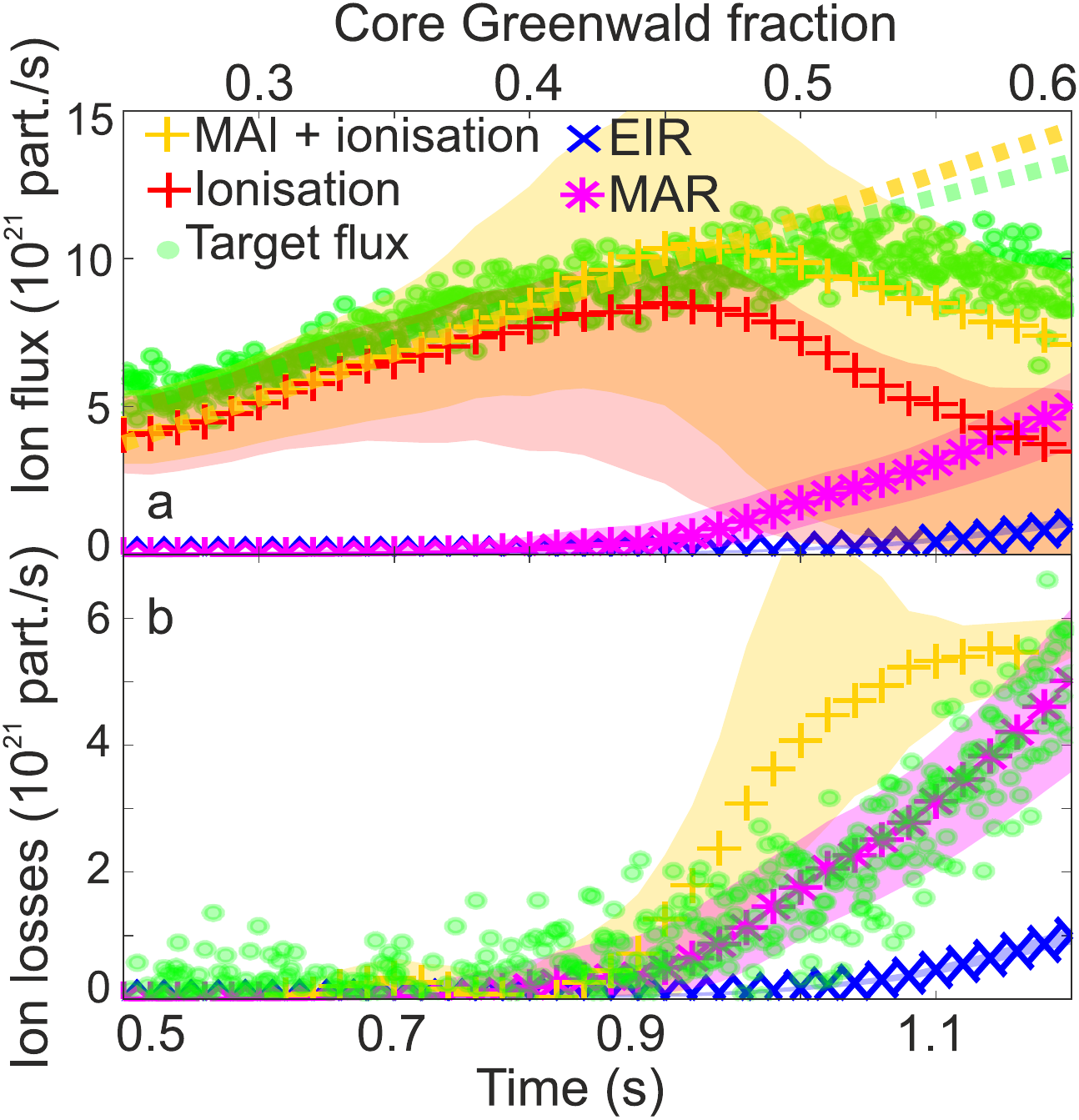}
    \caption{Inferred particle balance considering plasma-atom and plasma-molecule interactions. This modifies the estimated plasma-atom interaction processes (as it accounts for emission from plasma-molecule interactions in the spectroscopic analysis). a) Particle balance (ion target current) shown together with estimated atomic ion source, the sum of the atomic and MAI ($H_2^+$ and $H_2$) ion source, EIR sink and total ($H_2^+$ and $H^-$) MAR sink. We have added a linear fit in the attached phase (dotted lines) to the total ion source and ion target flux . b) Ion losses of the ion target flux, ion sources and through EIR and MAR.}    
    \label{fig:PartBal}
\end{figure}

Particle balance is shown from the perspective of both atomic and plasma-molecule interactions in figure \ref{fig:PartBal} for the discharge discussed throughout this paper. The difference in the ionisation estimate between figure \ref{fig:PartBal} and figure \ref{fig:HaMeasurement} is due to the self-consistent consideration of both plasma-atom ($H$ and $H^+$) and plasma-molecule interactions (involving $H_2$, $H_2^+$ and $H^-$) leading to excited atoms in figure \ref{fig:PartBal}. Both the ionisation estimates provide similar results until the detachment onset point ($n_e/n_{GW} \sim 0.43$). After the detachment onset, the estimate of ionisation including plasma-molecule interactions is reduced compared to that obtained when only plasma-atom interactions are included; the EIR ion sink estimate is unaltered. 

During detachment, plasma-molecule interactions start to contribute to $H\gamma, H\delta$ (larger impact for lower $n\rightarrow 2$ compared to higher $n\rightarrow2$ transitions) - see figure \ref{fig:SepaEmiss}. Not including molecular effects leads to an underestimate of the 'atomic' $H\delta/H\gamma$ ratio (e.g. higher-n/lower-n ratio), which makes Balmer line emission 'appear' to be more excitation rather than recombination dominated \cite{Verhaegh2019a}, leading to a potentially significant ionisation overestimate. We conclude that a self-consistent consideration of plasma-atom and plasma-molecule interactions can be important for inferring information on electron impact excitation (of $H$) emission (and thus ionisation, characteristic excitation temperature and radiated power from atomic excitation).

The analysis shown in figure \ref{fig:PartBal} indicates that MAR is significant at the detached roll-over phase. The onset of MAR occurs between the onset of power limitation (detachment onset) and the onset of EIR ion sinks. The inferred magnitude of MAR for this discharge is $\sim5$ times higher than the magnitude of EIR despite the emission of $H\gamma, H\delta$ (figure \ref{fig:SepaEmiss}) being EIR dominated. 

In this particular case, the MAR ion sink (magenta symbols in figure \ref{fig:PartBal}a) represents a significant fraction of the ion target flux (green symbols in figure \ref{fig:PartBal}a) ($51 \pm 15$ \%) and thus plays an important role in divertor particle balance.

Plasma-molecule interactions can also increase the ion target flux through MAI. The MAI rate calculated for this case is significant and starts to occur around the detachment onset. The MAI ion source in the detached phase is smaller than the MAR ion sink. These MAI estimates have larger uncertainties and are more sensitive to chordal integration effects as the MAI/$H\alpha$ photon ratio depends on the relative strength between molecular charge exchange and $H_2$ ionisation, which is strongly temperature dependent \cite{Verhaegh2020}. However, the MAI outcomes are anti-correlated with the atomic ionisation and thus the sum (which has lower uncertainties) of MAI and atomic ion sources is shown in figure \ref{fig:PartBal}.% The MAI and atomic ion sources have been combined as the individual uncertainties in those are anti-correlated, reducing the uncertainty. %\footnote{To determine whether a plasma-molecule reaction with $H_2^+$ leads to MAR or MAI one needs to identify the specific reaction which creates $H_2^+$: a) molecular charge exchange, which \emph{neutralises} $H^+$; or b) $H_2$ ionisation, which \emph{does not neutralise} $H^+$. We employ reaction rates from AMJUEL \cite{Reiter2008} to model the relative strength of the two $H_2^+$ creation rates and use these to assign MAR/MAI accordingly when computing the 'reaction to emission coefficient' ratios in the analysis chain.} 

Most MAI in this case arises from $H_2^+$ ions (formed from $H_2$ ionisation - $e^- + H_2 \rightarrow 2 e^- + H_2^+$) in a fairly high temperature regime ($T_e = [4-9]$ eV as opposed to $T_e = [1.5 - 4]$ eV for MAR) near the electron-impact excitation and Fulcher emission regions, which will be discussed in section \ref{ch:FulcherDiscus}. %Molecules likely reach this hot region during the detachment onset as the molecular density near the target rises and some of those molecules transport around the divertor leg radially into the relatively hot ionising plasma. 

To obtain a more quantitative comparison of the magnitude of ion losses through ion sinks, the reduction of the ion source and the reduction of the target ion flux, the 'ion losses' for these different processes are estimated, analogously to \cite{Verhaegh2019}, and are shown in figure \ref{fig:PartBal}b. We observe that ion source losses start to occur around the detachment onset and increase from that point onwards. The detachment process starts with the total ion source losses (e.g. including MAI), which seems to be the strongest (together with MAR) contributor to the ion target flux drop. This is followed by MAR and ultimately EIR ion sinks. The magnitude of ion loss due to MAR and the observed ion loss at the target are similar. 

The sum of the ion source and sink losses, however, exceeds the estimated ion target flux loss (figure \ref{fig:PartBal}) during detachment. This suggests the presence of an upstream ion flow towards the target ($\Gamma_u$ - equation \ref{eq:PartBal}) during detachment and thus a loss of 'high recycling conditions'. This contrasts previous findings \cite{Verhaegh2019} where only atomic reactions were considered and could arise from ionisation occurring above the spectrometer's divertor chordal view range (figure \ref{fig:HaMeasurement}d), which would be consistent with SOLPS-ITER simulations for TCV \cite{Fil2017,Fil2019submitted,Wensing2019}.

%\footnotetext{Two versions of the MAR sink are shown: a higher and lower estimate. If the total measured $H\alpha$ is equal to or below the atomic upscaled $H\alpha$, the associated MAR rate is zero. Having multiple zero estimates in the analysis sample of at a single time/spatial point implies a secondary peak in the probability density function near zero. For the higher MAR estimates these have been removed; this provides a more accurate estimate of MAR if MAR is present. The lower MAR estimates provide a more accurate indicator of when the analysis is confident that MAR is indeed present.}

\section{Discussion}
\label{ch:discussion}

\subsection{The evolution of plasma-molecule collisions and reactions during detachment}
\label{ch:FulcherDiscus}

In figure \ref{fig:FulcherBalmerSchem} we made the distinction between \emph{collisions} of the plasma and $H_2$, exciting $H_2$ rovibronically and \emph{reactions} between the plasma and $H_2^+$ (and possibly $H^-$) leading to excited atoms and Balmer line emission. The results from section \ref{ch:DetachEvolution} indicated that the Balmer line emission due to plasma-molecule interactions and Fulcher band emission emit at different locations and evolve differently during detachment. This suggests that there is a large volume in TCV of significant $H_2$ density extending over a range in $T_e$. Different kinds of plasma-molecule interactions dominate at different positions in this volume.

Fulcher band emission occurs when electrons have sufficient energy to electronically excite $H_2$ ($T_e > 4$ eV), which likely coincides with the region where $H_2$ is both dissociated as well as ionised \cite{Hollmann2006}. Therefore, Fulcher band emission should be fairly well localised around the hot part of the separatrix. It is thus surprising that Fulcher band emission occurs throughout the ionisation region given that the $H_2$ primary source is at/near the target and mean free paths at the target \emph{in attached conditions} are a few centimetres according to simulations \cite{Fil2017}. This suggests that molecules enter the divertor leg radially throughout the ionisation region, which could be attributed to the open, unbaffled divertor structure at the time. According to our analysis, those molecules penetrating into the ionisation region are responsible for the significant levels of MAI inferred (figure \ref{fig:PartBal}). The measurement that Fulcher emission occurs along the entire divertor leg and follows the ionisation region during detachment (figure \ref{fig:HaFulcherEmiss}) is qualitatively consistent with estimating the Fulcher emission profile by employing synthetic diagnostics on TCV SOLPS simulations from \cite{Fil2017}.%, which is spatially correlated with both the Fulcher emission and atomic ionisation regions. 

These findings seem to be in contrast with filtered camera imaging findings in DIII-D which indicate that 1) Fulcher emission is present \cite{Hollmann2006} in a thin layer close to the target; 2) the Fulcher emission region remains close towards the target during detachment. TCV has longer mean free paths for neutrals (5-10 cm near the ionisation region) and molecules than DIII-D which could contribute to this discrepancy (due to lower electron densities, lower heating powers as well as an open divertor). We measure the ionisation front ($T_e=[4-6]$ eV) is lifted $\sim$ 20 cm off the target during our experiment ($n_e/n_{GW} \sim 0.55$). Longer mean free paths and a larger distance between the target and ionisation front would facilitate a larger region in which plasma-molecule interactions occur. Additionally, the presence of other carbon emission lines \cite{Hollmann2006} in the bandpass filter during the measurement on DIII-D could also contribute to the Fulcher emission region not detaching from the target. 

In comparison to the primary location of Fulcher band emission, most Balmer line emission from $H_2^+$ (and possibly $H^-$) and thus the MAR ion sink as well as the $H_2^+$, $H^-$ densities, remains peaked near the target and extends up until the ionisation region. Near the target during detachment ($T_e = [1 - 3]$ eV) the electrons have insufficient energy to promote $H_2$ ionisation. However, \emph{vibrationally excited molecules}  still promote the formation of $H_2^+$ through molecular charge exchange. In other words, at the location where MAR occurs and most Balmer line emission due to $H_2$ chemistry is observed, \emph{vibrationally excited} molecules are likely responsible for the formation of $H_2^+$ (and possibly $H^-$), whereas at the Fulcher band emission region $H_2$ is \emph{electronically excited}, dissociated and ionised into $H_2^+$ resulting in MAI.     %The relation between the evolution of the Fulcher emission and the molecular contribution to the Balmer emission provides insight on the ordering of molecular process during detachment.   %This is schematically illustrated in figure \ref{fig:SchemEmiss}.

If there is no longer sufficient energy ($T_e < [4-5]$ eV) for the impact electrons to electronically excite $H_2$, this would result in strongly reduced Fulcher band emission. Our results indeed indicate that the Fulcher band emission is particularly dim below the ionisation region where most molecules (as well as MAR) are expected to be present. This may have implications for the applicability of using \emph{only} Fulcher band analysis to diagnose MAR. 
    %Note: The orange box wants to show $H_2^*(v)$ or $H_2^*(v)/H_2(v)$ as the Fulcher emission is coming from an electronic excited state (even if you have now vibrational excitation). Also $H_2^+$ doesn't require vibrational or electron excitation and can be produced from the ground state.}

% One may speculate that the results may indicate that the collisions between the plasma and the molecules near the ionisation front raise the density of the higher vibrational states locally. Those vibrationally excited molecules possibly transport throughout the molecular cloud; providing a sufficient amount of molecular excitation to promote $H_2^+$ and/or $H^-$ creation \emph{from the ionisation front up until the target}. This could be facilitated by the relatively large molecular dissociation mean free path below the ionisation region. This hypothesis could possibly explain why MAR remains peaked near the target as the $H_2$ density is highest near the target (due to the lower target temperatures). This could be exacerbated by relatively long-legged divertors, since such divertors could facilitate the formation of a large dense molecular cloud from the target to the ionisation region.

Although the mean-free-paths \emph{between the target and the ionisation region} for $H_2$ are fairly large \emph{in detached conditions} due to the lower electron temperatures, the expected mean free paths of $H_2^+$ and $H^-$, are much smaller. Potentially, transport of vibrationally excited molecules between the target and the ionisation front (and their interaction with the wall \cite{Fantz2002}) may play a role in achieving the higher vibrationally excited states near the target. Additionally, if electron impact collisions with $H_2$ have no longer enough energy to \emph{electronically excite} $H_2$ ($T_e < [4-5]$ eV) a larger proportion of the energy transfer during those collisions could go into raising the \emph{vibrational levels} in the electronic ground state, which is consistent with measurements and (vibrational-state resolved) simulations \cite{Fantz2001}.

\subsection{Evolution of hydrogenic line emission with plasma-molecule interactions and opacity}

\begin{figure}[H]
    \centering
    \includegraphics[width=\linewidth]{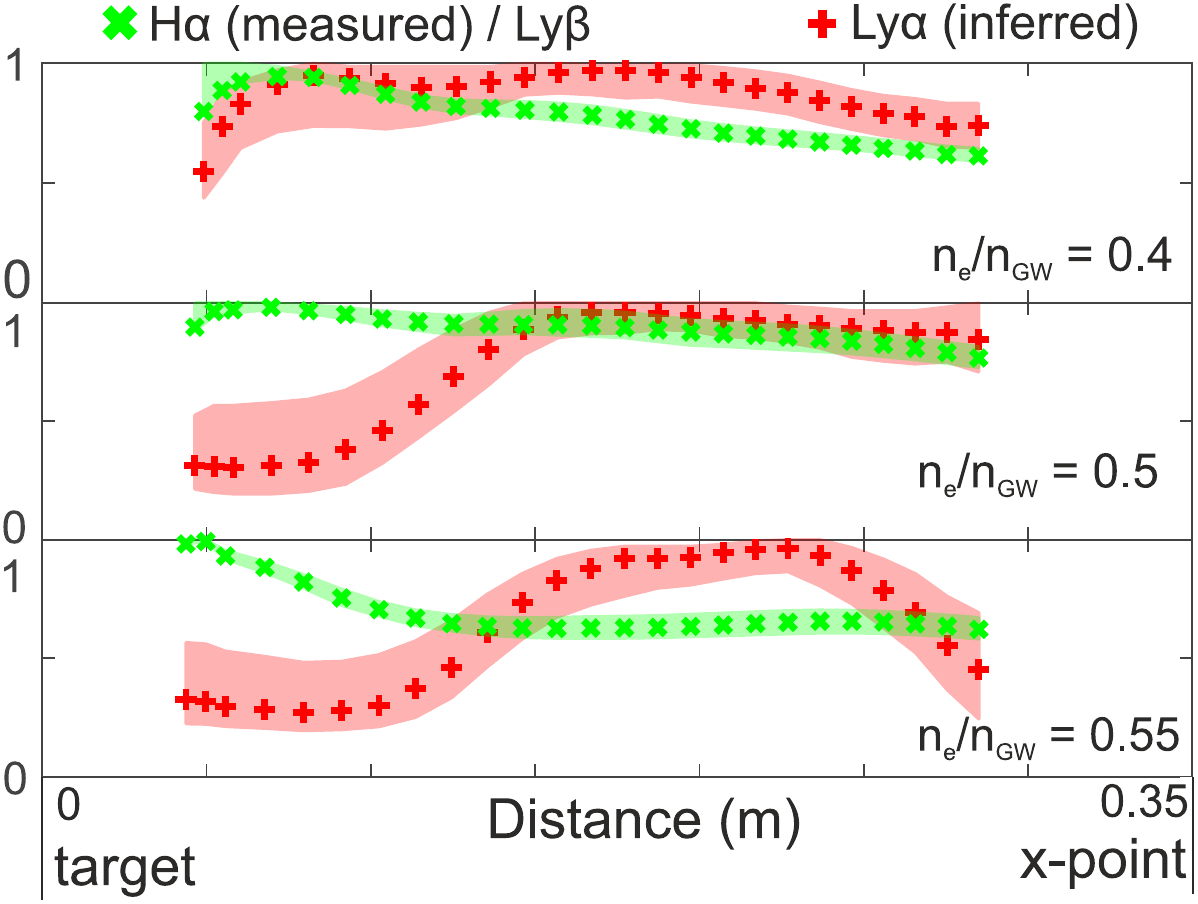}
    \caption{Normalised profiles (along the different spectroscopic chords) of the measured $H\alpha$ (and thus $Ly\beta$) emission and the inferred $Ly\alpha$ emission at three different core Greenwald fractions.}
    \label{fig:HaLyASepa}
\end{figure}

Our TCV results also indicate a spatial separation due to plasma-molecule interactions during detachment between the measured $H\alpha$ ($/ Ly\beta$ - $n=3$) and inferred (by extrapolating the individual emission processes) $Ly\alpha$ emission regions as shown in figure \ref{fig:HaLyASepa}. 

\begin{figure}[H]
    \centering
    \includegraphics[width=1\linewidth]{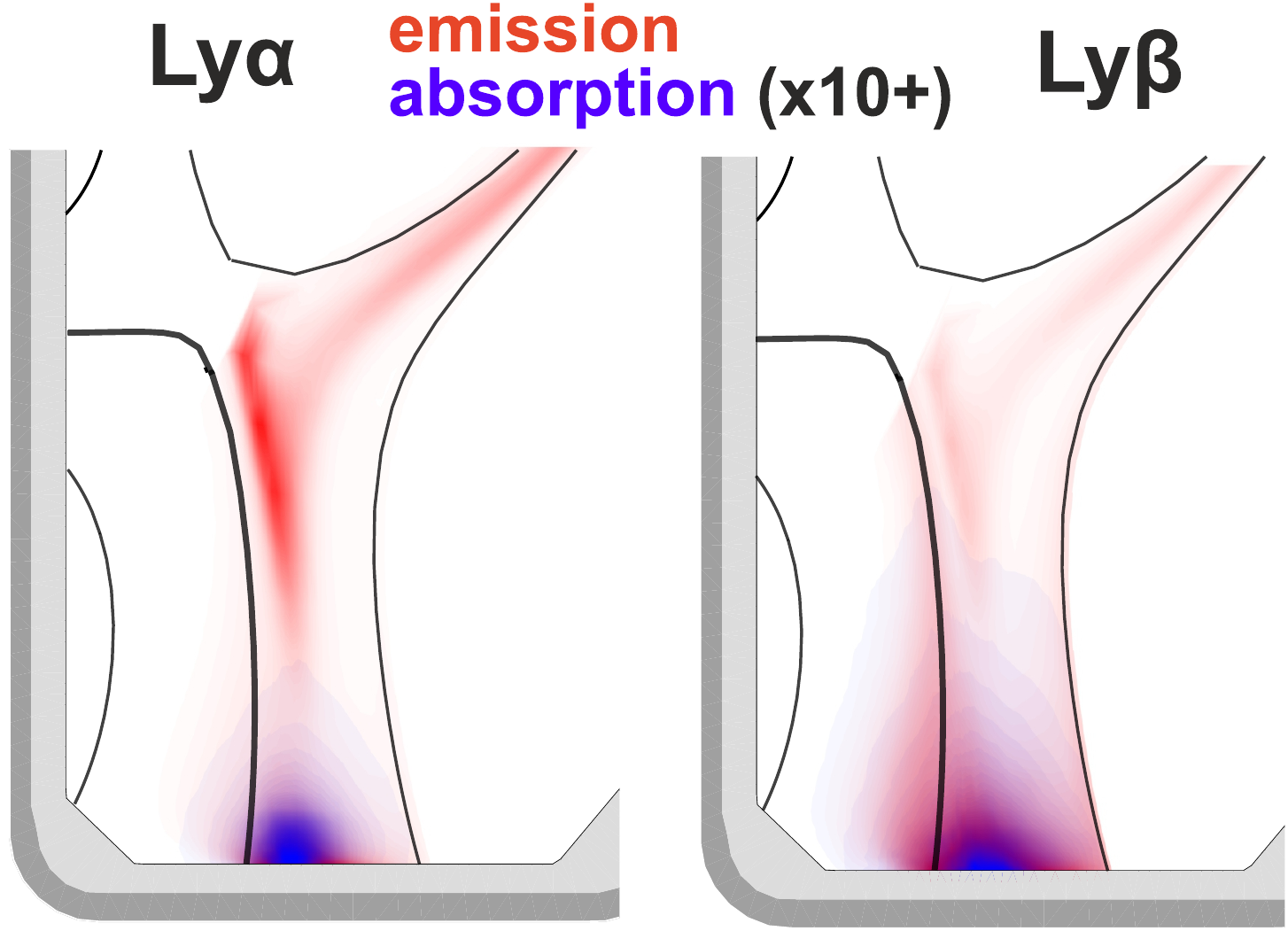}
    \caption{Schematic overview of $Ly\alpha$ and $Ly\beta$ absorption/emission based on ray-tracing obtained from post-processing a SOLPS-ITER simulation (\# 106278 \cite{Fil2017}) of a detached TCV discharge. As $Ly\alpha, Ly\beta$ opacity is negligible ($<5\%$), the magnitude of opacity is amplified in the figure for visibility. Emission/absorption from the inner target is omitted as the absorption there is not calculated to reduce computational time.}
    \label{fig:OpacitySchem}
\end{figure}

Figure \ref{fig:OpacitySchem} shows a TCV SOLPS simulation from \cite{Fil2017}, which is first post-processed to obtain Balmer line emissivities which include plasma-molecule interactions (section \ref{ch:analysis}) and further post-processed using ray-tracing to estimate where the emission gets absorbed due to opacity. This shows negligible levels of $Ly\alpha$ and $Ly\beta$ opacity is negligible on TCV ($<5\%$), hence we amplified the absorption in figure \ref{fig:OpacitySchem} for visibility. Plasma-molecule interactions enhance $H\alpha$ predominantly near the target resulting in a spatial separation of the $Ly\alpha$ and $Ly\beta$ emission regions, qualitatively consistent with figure \ref{fig:HaLyASepa}. As a result, $Ly\beta$ emits in a region where opacity is (relatively) more dominant than for $Ly\alpha$, due to the higher neutral densities near the target; significantly enhancing $Ly\beta$ opacity. %Additionally, the expected separation of the $Ly\alpha$ and $Ly\beta$ emission regions from the experimental analysis (figure \ref{fig:HaLyASepa}) is in qualitative agreement with that of the SOLPS 

Although photon opacity is expected to be insignificant for the TCV case studied here, opacity could play a stronger role in other devices such as JET, C-Mod and MAST-U where the photons traverse an integrated neutral density higher than $10^{18} m^{-2}$ \cite{Terry1998,Behringer2000,Lomanowski2020}. Post-processing MAST-U SOLPS-ITER simulations from \cite{Myatra} indicates significant opacity levels for $Ly\alpha$ and $Ly\beta$, whose emission regions are also separated resulting in a relatively higher opacity for $Ly\beta$. The separation of the $Ly\alpha$ and $Ly\beta$ emission regions could have implications for the diagnosis and understanding of (photon) opacity as using $Ly\beta$ opacity measurements to model the $Ly\alpha$ opacity (or the influence of opacity on the ionisation and recombination rates) \cite{Terry1998,Behringer2000,Lomanowski2020} requires assuming that $Ly\alpha$ and $Ly\beta$ emit and are opaque at the same regions along the line of sight. The development of spatially resolved Ly$\alpha$, Ly$\beta$ (as well as the Balmer series) diagnostics on tokamak divertors may be required.

\subsection{Implications for other devices}
\label{ch:ImplOtherDev}

This work highlights the importance of including plasma-molecule interactions in divertor plasma physics studies. There are several differences which could occur between these TCV results and other devices/conditions: %The fact that this analysis was performed on on TCV data, however, raises a concern as to the importance of plasma molecule interactions for detachment in other devices. 
\begin{enumerate}
    \item{Our research shows plasma-molecule interactions affect particle balance in a specific temperature regime $1<T_t<2.5$ eV - figure \ref{fig:HaMeasurement}. If the divertor remains detached while $T_t > 2.5$ eV, plasma-molecule interactions with $H_2^+$ (and possibly $H^-$) may not play a strong role. Furthermore, the importance of MAR is reduced in $T_t < 1$ eV detached conditions where 3-body recombination rises strongly and the EIR/MAR ratio is increased.} %We will investigate this further in future studies of other TCV detached discharges.
    \item{The transport of molecules and vibrational states in the molecular cloud may be different on TCV than other tokamaks due to differences in: 1) divertor geometry (e.g. baffled vs non-baffled, divertor chamber walls tight around the divertor leg); 2) wall material \cite{Fantz2002} (e.g. carbon vs tungsten); 3) molecular mean free paths (reduced with higher electron densities and heat fluxes). The transport of $H_2$ to higher temperature regions (sections \ref{ch:DetachEvolution} and \ref{ch:FulcherDiscus}) where MAI occurs may be related to the open TCV geometry and relatively large molecular mean free paths.}
    %\item{The transport of molecules (and what $T_e$ regions are accessed) will also be affected by changes in the mean free paths for molecular reactions driven by increases in either or both plasma density and heat fluxes. Higher electron densities and heat fluxes will reduce mean free paths of molecules. we expect that the balance of dominance between plasma-molecule and plasma-atom processes could change and the distance a molecule in any form will survive will shorten, potentially impacting the level of MAR.}
    \item{Higher electron densities than on TCV ($n_e\sim 10^{20} m^{-3}$) could increase the EIR/MAR ratio as electron-ion recombination scales with the power 2-3 of the electron density \cite{Terry1998,Verhaegh2017} while the plasma-molecule processes influencing MAR and MAI increase less quickly with electron density.}
    \item{The formation mechanisms of $H_2^+$ and $H^-$ are believed to be highly isotope dependent and could be different in deuterium-tritium plasmas.}
\end{enumerate}

Although differences may exist between TCV and conditions on other devices, experimental findings from both DIII-D \cite{Hollmann2006} and JET \cite{Lomanowski2020} using deuterium plasmas are also suggestive of plasma-molecule interactions influencing $H\alpha$. At DIII-D \cite{Hollmann2006}, the measured $H\alpha/H\beta$ line ratios were more than a factor 5 higher than that expected based on atomic interactions, which is in agreement with our TCV measurements. %At JET \cite{Lomanowski2020} the ion source inferred using the $H\alpha$ brightnesses (assuming that all $H\alpha$ emission was due to electron-impact excitation) increased strongly while the ion target current rolled-over; which could not be fully explained with electron-ion recombination or $Ly\beta$ opacity.  

Therefore, the importance of plasma-molecule interactions raised in this work may be generally applicable to other tokamaks with both deuterium and protium plasmas and further investigations on other devices are required.

\section{Summary}
\label{ch:conclusion} 

In this work we have applied new spectroscopic analysis techniques developed in \cite{Verhaegh2020} to investigate the impact of plasma-molecule interactions on the detachment process in TCV during core density ramp discharges. Multi-step plasma-molecule interaction processes produce $H_2^+$ (and possibly $H^-$) followed by the breakup of those species which leads to both Molecular Activated Recombination (MAR) as well as Molecular Activated Ionisation (MAI). Those reactions result in excited atoms, leading to enhancements of the Balmer series emission. We find that the impact of these reactions on both particle balance and the Balmer series emission is significant during detachment.

For our studied discharge, MAR results in $\sim 5$ times more ion loss than electron-ion recombination. While MAR is concentrated in the detached region it occurs at higher temperatures (up to 2.5 eV) than for 3-body recombination ($< 1$ eV). 

\emph{Balmer line emission} attributed to $H_2$ chemistry remains peaked at the target while \emph{$H_2$ Fulcher emission} 'detaches' from the target, following the ionisation region, as detachment proceeds. This suggests that there is a spatial separation between the various plasma-molecule interactions.

The strong enhancement of Balmer line emission near the target attributed to plasma-molecule interactions indicates enhancements in the Lyman series, potentially impacting power losses. These enhancements are expected to result in a separation of the $Ly\alpha$ and $Ly\beta$ emission regions which would likely lead to changes of the effects and location of opacity.

It appears that attributing the enhancements to Balmer series emission during detachment just to plasma-atom interactions can lead to an inaccurate description of divertor particle balance.

\section{Acknowledgements}

This work has received support from EPSRC Grant EP/T012250/1 and has been carriedout within the framework of the EUROfusion Consortium and has received funding from the Euratom research and training programme 2014-2018 and 2019-2020 under grant agreement No 633053.  The views and opinions expressed herein do not necessarily reflectthose of the European Commission.

\appendix

%Abc



%% References


\bibliographystyle{elsarticle-num}
\bibliography{all_bib.bib}

\begin{thebibliography}{10}
\expandafter\ifx\csname url\endcsname\relax
  \def\url#1{\texttt{#1}}\fi
\expandafter\ifx\csname urlprefix\endcsname\relax\def\urlprefix{URL }\fi
\expandafter\ifx\csname href\endcsname\relax
  \def\href#1#2{#2} \def\path#1{#1}\fi

\bibitem{Pitts2013}
R.~A. Pitts, S.~Carpentier, F.~Escourbiac, T.~Hirai, V.~Komarov, S.~Lisgo,
  A.~S. Kukushkin, A.~Loarte, M.~Merola, A.~S. Naik, R.~Mitteau, M.~Sugihara,
  B.~Bazylev, P.~C. Stangeby, A full tungsten divertor for {ITER}: Physics
  issues and design status, Journal of Nuclear Materials 438 (2013) S48--S56.
\newblock \href {https://doi.org/10.1016/j.jnucmat.2013.01.008}
  {\path{doi:10.1016/j.jnucmat.2013.01.008}}.

\bibitem{Verhaegh2020}
K.~Verhaegh, B.~Lipschultz, C.~Bowman, B.~P. Duval, U.~Fantz, A.~Fil, J.~R.
  Harrison, D.~Moulton, O.~Myatra, D.~Wünderlich, F.~Federici, D.~S. Gahle,
  A.~Perek, M.~Wensing, and, \href{https://doi.org/10.1088/1361-6587/abd4c0}{A
  novel hydrogenic spectroscopic technique for inferring the role of
  plasma{\textendash}molecule interaction on power and particle balance during
  detached conditions}, Plasma Physics and Controlled Fusion 63~(3) (2021)
  035018.
\newblock \href {https://doi.org/10.1088/1361-6587/abd4c0}
  {\path{doi:10.1088/1361-6587/abd4c0}}.
\newline\urlprefix\url{https://doi.org/10.1088/1361-6587/abd4c0}

\bibitem{Loarte2007}
A.~Loarte, B.~Lipschultz, A.~S. Kukushkin, G.~F. Matthews, P.~C. Stangeby,
  N.~Asakura, G.~F. Counsell, G.~Federici, A.~Kallenbach, K.~Krieger,
  A.~Mahdavi, V.~Philipps, D.~Reiter, J.~Roth, J.~Strachan, D.~Whyte,
  R.~Doerner, T.~Eich, W.~Fundamenski, A.~Herrmann, M.~Fenstermacher,
  P.~Ghendrih, M.~Groth, A.~Kirschner, S.~Konoshima, B.~LaBombard, P.~Lang,
  A.~W. Leonard, P.~Monier-Garbet, R.~Neu, H.~Pacher, B.~Pegourie, R.~A. Pitts,
  S.~Takamura, J.~Terry, E.~Tsitrone, I.~S.-o. L.~D. Phy, Chapter 4: Power and
  particle control, Nuclear Fusion 47~(6) (2007) S203--S263.
\newblock \href {https://doi.org/10.1088/0029-5515/47/6/S04}
  {\path{doi:10.1088/0029-5515/47/6/S04}}.

\bibitem{Goetz1996}
J.~A. Goetz, C.~Kurz, B.~LaBombard, B.~Lipschultz, A.~Niemczewski, G.~M.
  McCracken, J.~L. Terry, R.~L. Boivin, F.~Bombarda, P.~Bonoli, C.~Fiore,
  S.~Golovato, R.~Granetz, M.~Greenwald, S.~Horne, A.~Hubbard, I.~Hutchinson,
  J.~Irby, E.~Marmar, M.~Porkolab, J.~Rice, J.~Snipes, Y.~Takase, R.~Watterson,
  B.~Welch, S.~Wolfe, C.~Christensen, D.~Garnier, D.~Jablonski, D.~Lo,
  D.~Lumma, M.~May, A.~Mazurenko, R.~Nachtrieb, P.~OShea, J.~Reardon, J.~Rost,
  J.~Schachter, J.~Sorci, P.~Stek, M.~Umansky, Y.~Wang, Comparison of detached
  and radiative divertor operation in {Alcator C-Mod}, Physics of Plasmas 3~(5)
  (1996) 1908--1915.
\newblock \href {https://doi.org/10.1063/1.871986}
  {\path{doi:10.1063/1.871986}}.

\bibitem{LaBombard1997}
B.~LaBombard, J.~A. Goetz, I.~Hutchinson, D.~Jablonski, J.~Kesner, C.~Kurz,
  B.~Lipschultz, G.~M. McCracken, A.~Niemczewski, J.~Terry, A.~Allen, R.~L.
  Boivin, F.~Bombarda, P.~Bonoli, C.~Christensen, C.~Fiore, D.~Garnier,
  S.~Golovato, R.~Granetz, M.~Greenwald, S.~Horne, A.~Hubbard, J.~Irby, D.~Lo,
  D.~Lumma, E.~Marmar, M.~May, A.~Mazurenko, R.~Nachtrieb, H.~Ohkawa, P.~OShea,
  M.~Porkolab, J.~Reardon, J.~Rice, J.~Rost, J.~Schachter, J.~Snipes, J.~Sorci,
  P.~Stek, Y.~Takase, Y.~Wang, R.~Watterson, J.~Weaver, B.~Welch, S.~Wolfe,
  Experimental investigation of transport phenomena in the scrape-off layer and
  divertor, Journal of Nuclear Materials 241 (1997) 149--166.
\newblock \href {https://doi.org/10.1016/S0022-3115(96)00502-8}
  {\path{doi:10.1016/S0022-3115(96)00502-8}}.

\bibitem{Pitcher1997}
C.~S. Pitcher, P.~C. Stangeby, Experimental divertor physics, Plasma Physics
  and Controlled Fusion 39~(6) (1997) 779--930.
\newblock \href {https://doi.org/10.1088/0741-3335/39/6/001}
  {\path{doi:10.1088/0741-3335/39/6/001}}.

\bibitem{Labombard1995}
B.~Labombard, J.~Goetz, C.~Kurz, D.~Jablonski, B.~Lipschultz, G.~Mccracken,
  A.~Niemczewski, R.~L. Boivin, F.~Bombarda, C.~Christensen, S.~Fairfax,
  C.~Fiore, D.~Garnier, M.~Graf, S.~Golovato, R.~Granetz, M.~Greenwald,
  S.~Horne, A.~Hubbard, I.~Hutchinson, J.~Irby, J.~Kesner, T.~Luke, E.~Marmar,
  M.~May, P.~Oshea, M.~Porkolab, J.~Reardon, J.~Rice, J.~Schachter, J.~Snipes,
  P.~Stek, Y.~Takase, J.~Terry, G.~Tinios, R.~Watterson, B.~Welch, S.~Wolfe,
  {Scaling and Transport Analysis of Divertor Conditions on the Alcator C-Mod
  Tokamak}, Physics of Plasmas 2~(6) (1995) 2242--2248.
\newblock \href {https://doi.org/10.1063/1.871248}
  {\path{doi:10.1063/1.871248}}.

\bibitem{Stangeby2000}
P.~Stangeby, The plasma boundary of magnetic fusion devices, The Plasma
  Boundary of Magnetic Fusion Devices. Series: Series in Plasma Physics, ISBN:
  978-0-7503-0559-4. Taylor \& Francis, Edited by Peter Stangeby, vol. 7 7
  (2000).

\bibitem{Stangeby2018}
P.~C. Stangeby, Basic physical processes and reduced models for plasma
  detachment, Plasma Physics and Controlled Fusion 60~(4) (2018) 044022.

\bibitem{Verhaegh2019}
K.~Verhaegh, B.~Lipschultz, B.~Duval, O.~Février, A.~Fil, C.~Theiler,
  M.~Wensing, C.~Bowman, D.~Gahle, J.~Harrison, B.~Labit, C.~Marini,
  R.~Maurizio, H.~de~Oliveira, H.~Reimerdes, U.~Sheikh, C.~Tsui, N.~Vianello,
  W.~Vijvers", An improved understanding of the roles of atomic processes and
  power balance in divertor target ion current loss during detachment, Nuclear
  Fusion 59~(126038) (2019).
\newblock \href {https://doi.org/10.1088/1741-4326/ab4251}
  {\path{doi:10.1088/1741-4326/ab4251}}.

\bibitem{Lipschultz1999}
B.~Lipschultz, J.~L. Terry, C.~Boswell, J.~A. Goetz, A.~E. Hubbard, S.~I.
  Krasheninnikov, B.~LaBombard, D.~A. Pappas, C.~S. Pitcher, F.~Wising,
  S.~Wukitch, The role of particle sinks and sources in {Alcator C-Mod}
  detached divertor discharges, Physics of Plasmas 6~(5) (1999) 1907--1916.
\newblock \href {https://doi.org/10.1063/1.873448}
  {\path{doi:10.1063/1.873448}}.

\bibitem{Krasheninnikov2017}
S.~I. Krasheninnikov, A.~S. Kukushkin, Physics of ultimate detachment of a
  tokamak divertor plasma, Journal of Plasma Physics 83~(5) (2017) 155830501.
\newblock \href {https://doi.org/10.1017/S0022377817000654}
  {\path{doi:10.1017/S0022377817000654}}.

\bibitem{Pshenov2017}
A.~A. Pshenov, A.~S. Kukushkin, S.~I. Krasheninnikov, Energy balance in plasma
  detachment, Nuclear Materials and Energy 12 (2017) 948--952.
\newblock \href {https://doi.org/10.1016/j.nme.2017.03.019}
  {\path{doi:10.1016/j.nme.2017.03.019}}.

\bibitem{Stangeby2017}
P.~C. Stangeby, S.~Chaofeng, Strong correlation between {D 2} density and
  electron temperature at the target of divertors found in {SOLPS} analysis,
  Nuclear Fusion 57~(5) (2017) 056007.
\newblock \href {https://doi.org/10.1088/1741-4326/aa5e27}
  {\path{doi:10.1088/1741-4326/aa5e27}}.

\bibitem{Lomanowski2019}
B.~Lomanowski, M.~Carr, A.~Field, M.~Groth, S.~Henderson, J.~Harrison,
  A.~Huber, A.~Jarvinen, K.~Lawson, C.~Lowry, A.~Meigs, S.~Menmuir,
  M.~O’Mullane, M.~Reinke, C.~Stavrou, S.~Wiesen, Spectroscopic investigation
  of {N2} and {Ne} seeded induced detachment in {JET ITER}-like wall, Nuclear
  Materials and Energy 20~(100676) (2019,).
\newblock \href {https://doi.org/10.1016/j.nme.2019.100676}
  {\path{doi:10.1016/j.nme.2019.100676}}.

\bibitem{Fantz2002}
U.~Fantz, Emission spectroscopy of hydrogen molecules in technical and divertor
  plasmas, Contributions to Plasma Physics 42~(6-7) (2002) 675--684.
\newblock \href
  {https://doi.org/10.1002/1521-3986(200211)42:6/7<675::Aid-Ctpp675>3.0.Co;2-6}
  {\path{doi:10.1002/1521-3986(200211)42:6/7<675::Aid-Ctpp675>3.0.Co;2-6}}.

\bibitem{Fantz2001}
U.~Fantz, D.~Reiter, B.~Heger, D.~Coster, Hydrogen molecules in the divertor of
  {ASDEX Upgrade}, Journal of Nuclear Materials 290 (2001) 367--373.
\newblock \href {https://doi.org/10.1016/S0022-3115(00)00638-3}
  {\path{doi:10.1016/S0022-3115(00)00638-3}}.

\bibitem{Sakamoto2017}
M.~Sakamoto, A.~Terakado, K.~Nojiri, N.~Ezumi, Y.~Nakashima, K.~Sawada,
  K.~Ichimura, M.~Fukumoto, K.~Oki, K.~Shimizu, N.~Ohno, S.~Masuzaki, S.~Togo,
  J.~Kohagura, M.~Yoshikawa, Molecular activated recombination in divertor
  simulation plasma on {GAMMA 10/PDX}, Nuclear Materials and Energy 12 (2017)
  1004--1009.
\newblock \href {https://doi.org/10.1016/j.nme.2017.05.001}
  {\path{doi:10.1016/j.nme.2017.05.001}}.

\bibitem{Hollmann2006}
E.~M. Hollmann, S.~Brezinsek, N.~H. Brooks, M.~Groth, A.~G. McLean, A.~Y.
  Pigarov, D.~L. Rudakov, Spectroscopic measurement of atomic and molecular
  deuterium fluxes in the {DIII-D} plasma edge, Plasma Physics and Controlled
  Fusion 48~(8) (2006) 1165.

\bibitem{Groth2019}
M.~Groth, E.~Hollmann, A.~Jaervinen, A.~Leonard, A.~McLean, C.~Samuell,
  D.~Reiter, S.~Allen, P.~Boerner, S.~Brezinsek, I.~Bykov, G.~Corrigan,
  M.~Fenstermacher, D.~Harting, C.~Lasnier, B.~Lomanowski, M.~Makowski,
  M.~Shafer, H.~Wang, J.~Watkins, S.~Wiesen, R.~Wilcox, {EDGE2D-EIRENE}
  predictions of molecular emission in {DIII-D} high-recycling divertor
  plasmas, Nuclear Materials and Energy 19 (2019) 211--217.
\newblock \href {https://doi.org/10.1016/j.nme.2019.02.035}
  {\path{doi:10.1016/j.nme.2019.02.035}}.

\bibitem{Kukushkin2017}
A.~S. Kukushkin, S.~I. Krasheninnikov, A.~A. Pshenov, D.~Reiter, Role of
  molecular effects in divertor plasma recombination, Nuclear Materials and
  Energy 12 (2017) 984--988.
\newblock \href {https://doi.org/10.1016/j.nme.2016.12.030}
  {\path{doi:10.1016/j.nme.2016.12.030}}.

\bibitem{Wischmeier2004}
M.~Wischmeier, R.~A. Pitts, A.~Alfier, Y.~Andrebe, R.~Behn, D.~Coster,
  J.~Horacek, P.~Nielsen, R.~Pasqualotto, D.~Reiter, A.~Zabolotsky, The
  influence of molecular dynamics on divertor detachment in {TCV},
  Contributions to Plasma Physics 44~(1-3) (2004) 268--273.
\newblock \href {https://doi.org/10.1002/ctpp.200}
  {\path{doi:10.1002/ctpp.200}}.

\bibitem{Wuenderlich2016}
D.~W\"{u}nderlich, U.~Fantz, {Evaluation of State-Resolved Reaction
  Probabilities and Their Application in Population Models for He, H, and H2},
  Atoms 4~(4) (2016).
\newblock \href {https://doi.org/10.3390/atoms4040026}
  {\path{doi:10.3390/atoms4040026}}.

\bibitem{Wunderlich2020}
D.~W\"{u}nderlich, M.~Giacomin, R.~Ritz, U.~Fantz, Yacora on the web: Online
  collisional radiative models for plasmas containing h, h2 or he, Journal of
  Quantitative Spectroscopy and Radiative Transfer 240 (2020) 106695.
\newblock \href {https://doi.org/10.1016/j.jqsrt.2019.106695}
  {\path{doi:10.1016/j.jqsrt.2019.106695}}.

\bibitem{Kukushkin2016}
A.~Kukushkin, H.~Pacher, The role of “momentum removal” in divertor
  detachment, Contributions to Plasma Physics 56~(6‐8) (2016) 711--716.
\newblock \href {https://doi.org/10.1002/ctpp.201610048}
  {\path{doi:10.1002/ctpp.201610048}}.

\bibitem{Krishnakumar2011}
E.~Krishnakumar, S.~Denifl, I.~\ifmmode \check{C}\else
  \v{C}\fi{}ade\ifmmode~\check{z}\else \v{z}\fi{}, S.~Markelj, N.~J. Mason,
  Dissociative electron attachment cross sections for {${\mathrm{H}}_{2}$ and
  ${\mathrm{D}}_{2}$}, Phys. Rev. Lett. 106 (2011) 243201.
\newblock \href {https://doi.org/10.1103/PhysRevLett.106.243201}
  {\path{doi:10.1103/PhysRevLett.106.243201}}.

\bibitem{Lomanowski2020}
B.~Lomanowski, M.~Groth, I.~H. Coffey, J.~Karhunen, C.~F. Maggi, A.~Meigs,
  S.~Menmuir, M.~O'Mullane, {Interpretation of Lyman opacity measurements in
  JET with the ITER-like wall using a particle balance approach}, Plasma
  Physics and Controlled Fusion (2020).
\newblock \href {https://doi.org/10.1088/1361-6587/ab7432}
  {\path{doi:10.1088/1361-6587/ab7432}}.

\bibitem{Park2018}
J.~S. Park, M.~Groth, R.~Pitts, J.-G. Bak, S.~Thatipamula, J.-W. Juhn, S.-H.
  Hong, W.~Choe, Atomic processes leading to asymmetric divertor detachment in
  {KSTAR} {L-mode} plasmas, Nuclear Fusion 58~(12) (2018) 126033.
\newblock \href {https://doi.org/10.1088/1741-4326/aae865}
  {\path{doi:10.1088/1741-4326/aae865}}.

\bibitem{Myatra}
O.~Myatra, D.~Moulton, A.~Fil, B.~Dudson, B.~Lipschultz, {Taming the flame:
  Detachment access and control in MAST-U Super-X}, in: Plasma Surface
  Interactions, 2018.

\bibitem{Smolders}
A.~Smolders, M.~Wensing, S.~Carli, H.~D. Oliveira, W.~Dekeyser, B.~P. Duval,
  O.~F{\'{e}}vrier, D.~Gahle, L.~Martinelli, H.~Reimerdes, C.~Theiler,
  K.~Verhaegh, the TCV~team,
  \href{https://doi.org/10.1088/1361-6587/abbcc5}{Comparison of high density
  and nitrogen seeded detachment using {SOLPS}-{ITER} simulations of the
  tokamak {\'{a}} configuration variable}, Plasma Physics and Controlled Fusion
  62~(12) (2020) 125006.
\newblock \href {https://doi.org/10.1088/1361-6587/abbcc5}
  {\path{doi:10.1088/1361-6587/abbcc5}}.
\newline\urlprefix\url{https://doi.org/10.1088/1361-6587/abbcc5}

\bibitem{Moulton2018}
D.~Moulton, G.~Corrigan, J.~R. Harrison, B.~Lipschultz, Neutral pathways and
  heat flux widths in vertical- and horizontal-target {EDGE2D-EIRENE}
  simulations of {JET}, Nuclear Fusion 58~(9) (2018) 096029.
\newblock \href {https://doi.org/10.1088/1741-4326/aacf0f}
  {\path{doi:10.1088/1741-4326/aacf0f}}.

\bibitem{OMullane}
M.~O’Mullane, \href{http://www.adas.ac.uk}{{ADAS: Generalised Collisonal
  Radiative data for hydrogen}}, Tech. rep. (2013).
\newline\urlprefix\url{http://www.adas.ac.uk}

\bibitem{Summers2006}
H.~P. Summers, W.~J. Dickson, M.~G. O'Mullane, N.~R. Badnell, A.~D. Whiteford,
  D.~H. Brooks, J.~Lang, S.~D. Loch, D.~C. Griffin, Ionization state, excited
  populations and emission of impurities in dynamic finite density plasmas: I.
  the generalized collisional–radiative model for light elements, Plasma
  Physics and Controlled Fusion 48~(2) (2006) 263--293.
\newblock \href {https://doi.org/10.1088/0741-3335/48/2/007}
  {\path{doi:10.1088/0741-3335/48/2/007}}.

\bibitem{Reiter2005}
D.~Reiter, M.~Baelmans, P.~Börner, {The EIRENE and B2-EIRENE Codes}, Fusion
  Science and Technology 47~(2) (2005) 172--186.
\newblock \href {https://doi.org/10.13182/FST47-172}
  {\path{doi:10.13182/FST47-172}}.

\bibitem{Verhaegh2019a}
K.~Verhaegh, B.~Lipschultz, B.~Duval, A.~Fil, M.~Wensing, C.~Bowman, D.~Gahle,
  Novel inferences of ionisation \& recombination for particle/power balance
  during detached discharges using deuterium balmer line spectroscopy, Plasma
  Phys. Control. Fusion 61~(125018) (2019).
\newblock \href {https://doi.org/https://doi.org/10.1088/1361-6587/ab4f1e}
  {\path{doi:https://doi.org/10.1088/1361-6587/ab4f1e}}.

\bibitem{Reiter2008}
D.~Reiter, et~al., \href{http://www.eirene.de/manuals/eirene.pdf}{{The EIRENE
  code user manual}}, Report, Forschungszentrum Jülich GmbH (2008).
\newline\urlprefix\url{http://www.eirene.de/manuals/eirene.pdf}

\bibitem{Krasheninnikov1997}
S.~Krasheninnikov, A.~Y. Pigarov, D.~Knoll, B.~LaBombard, B.~Lipschultz,
  D.~Sigmar, T.~Soboleva, J.~Terry, F.~Wising, Plasma recombination and
  molecular effects in tokamak divertors and divertor simulators, Physics of
  Plasmas 4~(5) (1997) 1638--1646.

\bibitem{Fil2017}
A.~M.~D. Fil, B.~D. Dudson, B.~Lipschultz, D.~Moulton, K.~H.~A. Verhaegh,
  O.~Fevrier, M.~Wensing, Identification of the primary processes that lead to
  the drop in divertor target ion current at detachment in {TCV}, Contributions
  to plasma physics 58~(6-8) (2017).
\newblock \href {https://doi.org/10.1002/ctpp.201700171}
  {\path{doi:10.1002/ctpp.201700171}}.

\bibitem{Fil2019submitted}
A.~Fil, B.~Lipschultz, D.~Moulton, B.~D. Dudson, O.~F{\'{e}}vrier, O.~Myatra,
  C.~Theiler, K.~Verhaegh, M.~Wensing, and, Separating the roles of magnetic
  topology and neutral trapping in modifying the detachment threshold for
  {TCV}, Plasma Physics and Controlled Fusion 62~(3) (2020) 035008.
\newblock \href {https://doi.org/10.1088/1361-6587/ab69bb}
  {\path{doi:10.1088/1361-6587/ab69bb}}.

\bibitem{Wensing2019}
M.~Wensing, B.~Duval, O.~Fevrier, A.~Fil, D.~Galassi, E.~Havlickova, A.~Perek,
  H.~Reimerdes, C.~Theiler, K.~Verhaegh, M.~Wischmeier, {SOLPS-ITER simulations
  of the TCV divertor upgrade}, Plasma Phys. Control. Fusion 61~(085029)
  (2019).
\newblock \href {https://doi.org/10.1088/1361-6587/ab2b1f}
  {\path{doi:10.1088/1361-6587/ab2b1f}}.

\bibitem{Terry1998}
J.~L. Terry, B.~Lipschultz, A.~Y. Pigarov, S.~I. Krasheninnikov, B.~LaBombard,
  D.~Lumma, H.~Ohkawa, D.~Pappas, M.~Umansky, Volume recombination and opacity
  in {Alcator C-Mod} divertor plasmas, Physics of Plasmas 5~(5) (1998)
  1759--1766.
\newblock \href {https://doi.org/Doi 10.1063/1.872845} {\path{doi:Doi
  10.1063/1.872845}}.

\bibitem{Behringer2000}
K.~Behringer, U.~Fantz, The influence of opacity on hydrogen excited-state
  population and applications to low-temperature plasmas, New Journal of
  Physics 2 (2000) 23--23.
\newblock \href {https://doi.org/10.1088/1367-2630/2/1/323}
  {\path{doi:10.1088/1367-2630/2/1/323}}.

\bibitem{Verhaegh2017}
K.~Verhaegh, B.~Lipschultz, B.~P. Duval, R.~Harrison, H.~Reimerdes, C.~Theiler,
  B.~Labit, R.~Maurizio, C.~Marini, F.~Nespoli, U.~Sheikh, C.~K. Tsui,
  N.~Vianello, W.~A.~J. Vijvers, T.~T. . E.~M. Team, Spectroscopic
  investigations of divertor detachment in {TCV}, Nuclear Materials and Energy
  12 (2017) 1112--1117.
\newblock \href {https://doi.org/10.1016/j.nme.2017.01.004}
  {\path{doi:10.1016/j.nme.2017.01.004}}.

\end{thebibliography}






% \end{thebibliography}

\end{document}